\documentstyle[12pt,epsfig]{article}
\date{\today}

\newcommand{\scs}{\scriptscriptstyle}
\newcommand{\be}{\begin{equation}}
\newcommand{\ee}{\end{equation}}
\newcommand{\bea}{\begin{eqnarray}}
\newcommand{\eea}{\end{eqnarray}}
\newcommand{\f}{\frac}
\def\slash#1{\setbox0=\hbox{$#1$}#1\hskip-\wd0\dimen0=5pt\advance
       \dimen0 by-\ht0\advance\dimen0 by\dp0\lower0.5\dimen0\hbox
         to\wd0{\hss\sl/\/\hss}}

\title{Single top quark production via SUSY-QCD FCNC couplings at the CERN LHC in the unconstrained MSSM}
\author{Jian Jun Liu$^a$, Chong Sheng Li$^a$\footnote{csli@pku.edu.cn}, Li Lin Yang$^a$ and Li Gang Jin$^b$\\
{\small $^a$ Department of Physics, Peking University, Beijing
100871, China} \\ {\small $^b$ Institute of Theoretical Physics,
Academia Sinica, P. O. Box 2735,
Beijing 100080, China} \\
}
\date{\today}

\begin{document}
\maketitle
\begin{abstract}
We evaluate the $t\bar{c}$ and $t\bar{u}$ productions at the LHC
within the general unconstrained MSSM framework. We find that
these single top quark productions induced by SUSY-QCD FCNC
couplings have remarkable cross sections for favorable parameter
values allowed by current low energy data, which can be as large
as a few pb. Once large rates of the $t\bar{c}$ and $t\bar{u}$
productions are detected at the LHC, they may be induced by SUSY
FCNC couplings. We show that the precise measurement of single top
quark production cross sections at the LHC is a powerful probe for
the details of the SUSY FCNC couplings.
\end{abstract}

\vspace{1.5cm} \noindent  Keywords: top quark, MSSM, FCNC


\newpage

\section{Introduction}
The flavor dynamics, such as the mixing of three generation
fermions and their large mass differences observed, is still a
great mystery in particle physics today. The heaviest one of three
generation fermions is the top quark with the mass close to the
electroweak (EW) symmetry breaking scale. Therefore, it can play a
role of a wonderful probe for the EW breaking mechanism and new
physics beyond the standard model (SM) through its decays and
productions. An important aspect of the top quark physics is to
investigate anomalous flavor changing neutral current (FCNC)
couplings. Within the SM, FCNC is forbidden at the tree-level and
highly suppressed by the GIM mechanism \cite{GIM} at one-loop
level. All the precise measurements of various FCNC processes, for
example, $b \rightarrow s\gamma$ \cite{pdg}, agree with the SM
predictions. However, many new physics models allow the existence
of the tree-level FCNC couplings, and may enhance some FCNC
processes. In particular, the minimal supersymmetric standard
model (MSSM) is one of the most popular new physics models. And it
is hard to believe that this model will say no more about the
flavor dynamics than the SM. The MSSM includes 105 free parameters
beyond the SM: 5 real parameters and 3 CP-violating phases in the
gaugino/higgsino sector, 21 squark and slepton masses, 36 new real
mixing angles to define the squark and slepton mass eigenstates,
and 40 new CP-violating phases that can appear in squark and
slepton interactions \cite{pdg}, so in general, it can provide a
new explanation of the source of FCNC couplings. Actually, many of
the parameters are constrained by present experimental data, and
some popular SUSY models, for example, the gravity-mediated
supersymmetry broken model (SUGRA) \cite{sugra}, gauge mediated
supersymmetry broken model (GMSB) \cite{gmsb}, anomaly mediated
supersymmetry broken model (AMSB) \cite{amsb}, gaugino mediated
supersymmetry broken model (gMSB) \cite{gmsb1} and Kaluza-Klein
mediated supersymmetry broken model \cite{kksb}, make some ad hoc
assumptions to reduce the number of independent parameters.
Although the predictions of these models can agree with the
present constraints from FCNC experiments, it should be noted that
few experimental constraints on the top quark FCNC processes are
available so far \cite{pdg}. Since the CERN Large Hadron Collider
(LHC) can produce abundant top events, the measurement of the top
quark rare processes will become possible. Therefore, a good
understanding of the theoretical predictions of these processes,
especially within the general unconstrained MSSM framework, is
important. And a careful investigation will provide some clues to
the constraint relations among the supersymmetry (SUSY) parameters
and more clear information about the SUSY breaking mechanisms.

The top quark FCNC processes have been investigated in detail.
There are two categories of these investigations, one of which is
the top quark FCNC decays \cite{sola}, especially for $t
\rightarrow cg$ \cite{csldecay,plb} and $t \rightarrow ch$
\cite{csldecayh} in the MSSM. These results show that the branch
ratio of $t \rightarrow cg$ can reach $10^{-4}$ \cite{plb}, which
enhances the one in the SM ($\sim 10^{-11}$) significantly. The
other category is the top quark productions through FCNC processes
\cite{tait}--\cite{csli}, ones of which induced by the anomalous
top quark FCNC couplings at the colliders have been extensively
discussed in a model independent way \cite{tait}--\cite{gouz} and
model dependent way \cite{csli}, respectively. As we know, most of
the works within SUSY models are limited in some constrained MSSM,
where the top quark FCNC couplings were studied through some
SUSY-CKM matrices or mass insertion approximation  \cite{hall}.
However, it is interesting to study the top quark FCNC productions
using soft SUSY broken parameters directly, and then establish
some relations between the production cross sections and soft SUSY
broken parameters in the mass eigenstate formalism
\cite{besmer,harnik}. Actually, the general framework with SUSY
FCNC mechanisms was presented several years ago \cite{rosiek}, and
there have been a lot of works studying SUSY FCNC processes within
this framework \cite{besmer,harnik}. But the top quark SUSY FCNC
production processes have not been studied in above framework so
far, and this means that the relations between the observables of
the top quark production processes and the mixing among up squarks
involving the third generation have not been established yet. In
this paper, we will investigate the single top quark productions
at the LHC induced by SUSY FCNC couplings in the general
unconstrained MSSM, which include the top quark productions
associated with the anti-charm quark and the anti-up quark, and we
believe that it is an important step towards revealing the top
quark SUSY FCNC couplings and exploring some relations among SUSY
broken parameters at the LHC.

The paper is organized as follows: In Sec. \ref{sec:cal} we
describe the general framework of SUSY FCNC mechanisms. In Sec.
\ref{sec:pro} we evaluate the total production cross sections of
the $pp\rightarrow t\bar{c}(\bar{u})$ process induced by SUSY FCNC
couplings, and the detail expressions for various couplings and
form factors are given in the Appendices A and B, respectively. In
Sec. \ref{sec:dis} we present our numerical calculations and
discussions. Finally, Sec. \ref{sec:con} gives our conclusions.

\section{\label{sec:cal}The FCNC in the MSSM}

In order to make our paper self-contained, we start with a brief
description of the FCNC mechanism in the MSSM, as shown in
Ref.~\cite{rosiek}, and establish our notation conventions.

The SUSY part of the MSSM Lagrangian is
\bea \label{eq:Lsusy} {\cal L}_{\scs SUSY} &=& -\f{1}{4}
F_G^{a\;\mu\nu} F^a_{G\;\mu\nu} + i \bar{\lambda}^a_G {\slash
D}_{ab} \lambda^b_G + (D^{\mu}\phi)^{\dagger} (D_{\mu}\phi) + i
\bar{\psi} {\slash D} \psi
\nonumber\\
&& - \left( \f{ \partial \tilde{\mathcal{W}}}{\partial \phi_i}
\right)^{\star}
  \left( \f{ \partial \tilde{\mathcal{W}}}{\partial \phi_i} \right)
-\f{1}{2} \left( \f{ \partial^2 \tilde{\mathcal{W}}}{\partial
\phi_i
\partial \phi_j} \psi_i^T C \psi_j \; + \; {\rm h.c.} \right) \nonumber\\
&& -\sqrt{2} g_G \left( \phi^{\dagger} T^a_G \lambda^{a\;T}_G C
\psi \; + \; {\rm h.c.} \right) -\f{1}{2} g_G^2 (\phi^{\dagger}
T^a_G \phi)(\phi^{\dagger} T^a_G \phi), \eea
where
\be \label{eq:W} \tilde{\mathcal{W}} = \mu H^1 H^2 + Y_l^{IJ} H^1
\tilde{L}^I \tilde{E}_R^J + Y_d^{IJ} H^1 \tilde{Q}^I \tilde{D}_R^J
+ Y_u^{IJ} H^2 \tilde{Q}^I \tilde{U}_R^J. \label{eq:superpot} \ee
Here the subscript $G$ represents color, weak isospin and
supercharge indices of the SM gauge group $SU(3)_C \times SU(2)_L
\times U(1)_Y$, respectively, $a$ and $b$ are the indices of
adjoint representations of the non-abelian subgroups, and $I, J =
1, 2, 3$ are the generation indices. $\phi$, $\psi$ and $\lambda$
represent scalar fields, matter fermion fields and gauginos,
respectively. $C$ is charge conjugation matrix. $\tilde{L}^I(\tilde{Q}^I)$
and $\tilde{E}_R^I(\tilde{U}_R^I,\tilde{D}_R^I)$ represent
slepton(squark) $SU(2)$ left-hand doublets and right-hand
singlets, respectively. $H^{1,2}$ represent two Higgs $SU(2)$ doublets,
and their vacuum expectation values are
\be \langle H^1 \rangle = \left( \begin{array}{c}
\f{v_1}{\sqrt{2}} \\ 0 \end{array} \right) \equiv \left(
\begin{array}{c} \f{v \cos \beta}{\sqrt{2}} \\ 0 \end{array}
\right), \hspace{2cm} \langle H^2 \rangle = \left(
\begin{array}{c} 0 \\ \f{v_2}{\sqrt{2}} \end{array} \right) \equiv
\left( \begin{array}{c} 0 \\ \f{v \sin \beta}{\sqrt{2}}
\end{array} \right), \ee
where $v=(\sqrt{2}G_F)^{-1/2}=246$~GeV, and the angle $\beta$ is
defined by $\tan \beta\equiv v_2/v_1$, the ratio of the vacuum
expectation values of the two Higgs doublets. $\mu$ is the Higgs
mixing parameter.

The soft SUSY breaking Lagrangian is
\bea {\cal L}_{soft} &=& -\f{1}{2} \left(   M_3 \tilde{g}^{a\;T} C
\tilde{g}^a + M_2 \tilde{W}^{i\;T} C \tilde{W}^i + M_1 \tilde{B}^T
C \tilde{B} \; + \; {\rm h.c.} \right)\nonumber \\ && - M_{H^1}^2
H^{1\;\dagger} H^1 - M_{H^2}^2 H^{2\;\dagger} H^2 - \tilde{L}^{I
\dagger} ( M_L^2 )^{IJ} \tilde{L}^J - \tilde{E}_R^{I \dagger} (
M_E^2 )^{IJ} \tilde{E}_R^J \nonumber \\ && - \tilde{Q}^{I \dagger}
( M_Q^2 )^{IJ} \tilde{Q}^J - \tilde{D}_R^{I
\dagger} ( M_D^2 )^{IJ} \tilde{D}_R^J - \tilde{U}_R^{I \dagger} ( M_U^2 )^{IJ} \tilde{U}_R^J \nonumber \\
&& + \left( A_E^{IJ} H^1 \tilde{L}^I \tilde{E}_R^J
         + A_D^{IJ} H^1 \tilde{Q}^I \tilde{D}_R^J
         + A_U^{IJ} H^2 \tilde{Q}^I \tilde{U}_R^J
     + B \mu H^1 H^2 \; + \; {\rm h.c.} \right). \hspace{0.5cm}
\label{eq:lsoft} \eea
Here $M^2_Q$, $M^{2}_{U,D}$ and $A_{U,D}$ are
the soft broken $SU(2)$ doublet squark mass squared matrix, the
$SU(2)$ singlet squark mass squared matrix and the trilinear
coupling matrix, respectively.

Using $3 \times 3$ unitary matrices $V^{E,U,D}_{L,R}$, the lepton
and quark mass eigenstates are given by
\bea
\begin{array}{rclcrcl}
\nu &=&  V^E_L \Psi_{{\scs L}1}, & \hspace{2cm} & u   &=&  V^U_L
\Psi_{{\scs Q}1} \; + \; V^U_R C \bar{\Psi}_{\scs U}^T,
\vspace{0.2cm}\\
l &=&  V^E_L \Psi_{{\scs L}2} \; + \; V^E_R C \bar{\Psi}_{\scs
E}^T, & \hspace{2cm} & d   &=&  V^D_L \Psi_{{\scs Q}2} \; + \;
V^D_R C \bar{\Psi}_{\scs D}^T.
\end{array}
\eea
Their corresponding diagonal $3 \times 3$ mass matrices are
\bea m_l = -\f{v \cos \beta}{\sqrt{2}} V^E_R Y_l^T
V^{E\;\dagger}_L, \hspace{0.5cm} m_u =  \f{v \sin \beta}{\sqrt{2}}
V^U_R Y_u^T V^{U\;\dagger}_L, \hspace{0.5cm} m_d = -\f{v \cos
\beta}{\sqrt{2}} V^D_R Y_d^T V^{D\;\dagger}_L. \nonumber \eea
As in the SM, the Cabbibo-Kobayashi-Maskawa (CKM) matrix is $K =
V^U_L V^{D\;\dagger}_L$.

It is convenient to specify the squark mass matrices in the
super-CKM basis, in which the mass matrices of the quark fields
are diagonalized by rotating the superfields. The super-CKM basis
$\tilde{U}^0$ is defined as
\bea \tilde{U}^0 = \left( \begin{array}{c} V^U_L \tilde U_L \\
V^U_R \tilde U^{\ast}_R
\end{array} \right). \label{eq:superkm} \eea
And in this basis, the up squark mass
matrix is $6 \times 6$ matrix, which has the form:
\begin{equation}
{\cal M}_{\tilde{U}}^2 \equiv  \left( \begin{array}{cc}
  \left(M^2_{\,{\tilde{U}}}\right)_{LL} +F_{u\,LL} +D_{u\,LL}           &
  \left(M^2_{\,{\tilde{U}}}\right)_{LR} + F_{u\,LR}
                                                     \\[1.01ex]
 \left(M^2_{\,{\tilde{U}}}\right)_{LR}^{\dagger} + (F_{u\,LR})^\dagger &
             \ \ \left(M^2_{\,{\tilde{U}}}\right)_{RR} + F_{u\,RR} +D_{u\,RR}
 \end{array} \right) \,.
\label{massmatrixu}
\end{equation}
The $F$ terms  and $D$ terms are diagonal $3 \times 3$ submatrices, which are given by
\begin{equation}
 F_{u\,LR} =  -\mu (m_{u} \cot \beta)\,  {{\mathchoice
{\rm 1\mskip-4mu l} {\rm 1\mskip-4mu l} {\rm 1\mskip-4.5mu l} {\rm
1\mskip-5mu l}}}_3 \, , \qquad
F_{u\,LL} =  F_{u\,RR} = m^2_{u}\,
{{\mathchoice {\rm 1\mskip-4mu l} {\rm 1\mskip-4mu l} {\rm
1\mskip-4.5mu l} {\rm 1\mskip-5mu l}}}_3 \, , \label{Fterm}
\end{equation}
\begin{equation}
 D_{u\,LL} = D_{u\,RR} =  \cos 2\beta \, m_Z^2
   \left(\frac{1}{2} - \frac{2}{3} \sin^2\theta_W \right)
{{\mathchoice {\rm 1\mskip-4mu l} {\rm 1\mskip-4mu l} {\rm
1\mskip-4.5mu l} {\rm 1\mskip-5mu l}}}_3\,, \label{dterm}
\end{equation}
where $\theta_W$ is the Weinberg angle, ${{\mathchoice {\rm
1\mskip-4mu l} {\rm 1\mskip-4mu l} {\rm 1\mskip-4.5mu l} {\rm
1\mskip-5mu l}}}_3$ stands for the $3 \times 3$ unit matrix. And
$(M^2_{\tilde{U}})_{LL}$, $(M^2_{\tilde{U}})_{RR}$ and
$(M^2_{\tilde{U}})_{LR}$ contain the flavor-changing entries \bea
\begin{array}{ccc}
(M^2_{\tilde{U}})_{LL} = V_L^U M^2_Q V_L^{U\dagger},
\hspace{0.4cm}& (M^2_{\tilde{U}})_{RR} = V_R^U (M^{2}_U)^T
V_R^{U\dagger}, \hspace{0.4cm}& (M^2_{\tilde{U}})_{LR} = -\f{v
\sin \beta}{\sqrt{2}} V_L^U A_U^{\ast} V_R^{U\dagger},
\end{array}
\eea which are directly related to the mechanism of SUSY breaking,
and are in general not diagonal in the super-CKM basis.
Furthermore, $(M^2_{\tilde{U}})_{LR}$, arising from the trilinear
terms in the soft potential, is not hermitian.
 The matrix ${\cal M}^2_{\tilde{U}}$ can
further be diagonalized by an additional $6\times 6$ unitary
matrix $Z_U$ to give the up squark mass eigenvalues
\bea \left({\cal M}^2_{\tilde{U}}\right)^{diag} = Z_U^{\dagger}
{\cal M}^2_{\tilde{U}} Z_U \label{eq:zudef}. \eea
%

As for the down squark mass matrix, we also can define
${\mathcal{M}}_{\tilde{D}}^2$ as the similar form of
Eq.~(\ref{massmatrixu}) with the replacement of
$(M^2_{\tilde{U}})_{IJ}$ ($I,J=L,R$) by $(M^2_{\tilde{D}})_{IJ}$.
Note that since $SU(2)_L$ gauge invariance implies that
$(M^2_{\tilde{U}})_{LL} = K (M^2_{\tilde{D}})_{LL} K^\dagger$, the
matrices $(M^2_{\tilde{U}})_{LL}$ and $(M^2_{\tilde{D}})_{LL}$ are
correlated and cannot be specified independently.

Thus, in the super-CKM basis, there are new potential sources of
flavor-changing neutral current: neutralino-quark-squark coupling
and gluino-quark-squark coupling, which arise from the
off-diagonal elements of $(M^2_{\tilde{U}})_{LL}$,
$(M^2_{\tilde{U}})_{LR}$ and $(M^2_{\tilde{U}})_{RR}$. As the
former coupling is in general much weaker than the latter one, in
this paper, we only consider the SUSY-QCD FCNC effects induced by
the gluino-quark-squark coupling, which can be written as
\bea i\sqrt{2}g_sT^a \left[-(Z_{U})_{Ii}P_L + (Z_{U})_{(I+3)i}
P_R\right] \qquad \quad I,i=1,2,3. \eea
Here $P_{L,R}\equiv (1\mp \gamma_5)/2$ and $T^a$ is the $SU(3)$
color matrix. Thus the flavor changing effects of soft broken
terms $M^2_{Q}$, $M^2_{U}$ and $A_{U}$ on the observables can be
obtained through the matrix $Z_U$.

As mentioned in the introduction, most of previous literatures
studied the FCNC processes in the MSSM by so-called mass insertion
approximation \cite{hall}. However, when the off-diagonal elements
in the up squark mass matrix become large, the mass insertion
approximation is no longer valid \cite{besmer,harnik}. Therefore,
we use the general mass eigenstate formalism as described above,
which has been adopted in the literatures recently
\cite{plb,besmer,harnik}. So any effects of SUSY FCNC couplings in
loops on various observables will provide some information of the
SUSY breaking mechanism.

In practical calculations, because the squark mass matrix  ${\cal
M}^2_{\tilde{U}}$ is a $6\times6$ matrix, it is very formidable to
calculate the effects arising from the matrix on the observables
of FCNC processes. Thus, as shown in
Ref.~\cite{plb,besmer,harnik}, it is reasonable to calculate the
observables by only switching on {\it one} off-diagonal element in
$(M^2_{\tilde{U}})_{LL}$, $(M^2_{\tilde{U}})_{LR}$ and
$(M^2_{\tilde{U}})_{RR}$ at a time. For the aim of this paper, the
following strategy in the calculations of the process $pp
\rightarrow t\bar{c}(\bar{u})$ will be used: first we deal with
the LL, LR, RL and RR block of the matrix ${\cal M}^2_{\tilde{U}}$
separately and in each block we only consider the effects of
individual element on the production cross sections, and then we
investigate both the interference effects between different
entries within one block and the interference effects between
different blocks. To simplify the calculation we further assume
that all diagonal entries in $(M^2_{\tilde{U}})_{LL}$,
$(M^2_{\tilde{U}})_{LR}$, $(M^2_{\tilde{U}})_{RL}$ and
$(M^2_{\tilde{U}})_{RR}$ are set to be equal to the common value
$M^2_{\rm{SUSY}}$, and then we normalize the off-diagonal elements
to $M^2_{\rm{SUSY}}$ \cite{besmer,gam},
\begin{eqnarray}
&& (\delta_{U}^{ij})_{LL} =
\frac{(M^2_{\tilde{U}})_{LL}^{ij}}{M^2_{\rm{SUSY}}}\,,
\hspace{1.0truecm} (\delta_{U}^{ij})_{RR} =
\frac{(M^2_{\tilde{U}})_{RR}^{ij}}{M^2_{\rm{SUSY}}}\,,
\hspace{1.0truecm} \nonumber \\
&& (\delta_{U}^{ij})_{LR} =
\frac{(M^2_{\tilde{U}})_{LR}^{ij}}{M^2_{\rm{SUSY}}}\,,
\hspace{1.0truecm} (\delta_{U}^{ij})_{RL} =
\frac{(M^2_{\tilde{U}})_{RL}^{ij}}{M^2_{\rm{SUSY}}}\,,
\hspace{1.0truecm} (i \ne j,i,j=1,2,3). \label{deltadefb}
\end{eqnarray}
Thus $(M^2_{\tilde{U}})_{LL}$ can be written as follows: \be
(M^2_{\tilde{U}})_{LL} =M^2_{\rm SUSY} \left(
\begin{array}{ccc}
  1    & (\delta_{U}^{12})_{LL} & (\delta_{U}^{13})_{LL}
\vspace{0.2cm} \\
(\delta_{U}^{21})_{LL} &    1   & (\delta_{U}^{23})_{LL}
\vspace{0.2cm} \\
(\delta_{U}^{31})_{LL} & (\delta_{U}^{32})_{LL} & 1
\end{array}\right),
\ee
and analogously for all the other blocks.

\section{\label{sec:pro}The process $pp\rightarrow t\bar{c}(\bar{u})$ induced by
SUSY FCNC couplings} The process $pp\rightarrow t\bar{c}(\bar{u})
$ at the LHC can be induced through the SUSY-QCD FCNC couplings
with the initial partonic states $u\bar{u}$, $d\bar{d}$,
$s\bar{s}$, $c\bar{c}$, $u\bar{c}$, $c\bar{u}$ and $gg$, as shown
in Fig.1. Our practical calculations show that the contributions
from the $b\bar{b}$, $b\bar{s}$ and $s\bar{b}$ initial states are
negligibly small, so we do not discuss them below. As the $t
\bar{u}$ process is completely similar to the $t \bar{c}$ one,
here we only discuss the situations of $t \bar{c}$ in detail.
Neglecting the charm quark mass, the amplitude of the process $gg
\rightarrow t\bar{c}$ can be written as
\begin{eqnarray}
{\cal M}^{gg}={\cal M}^{gg}_{self}+{\cal M}^{gg}_{vertex}+{\cal M}^{gg}_{box},
\end{eqnarray}
where ${\cal M}^{gg}_{self}$, ${\cal M}^{gg}_{vertex}$ and ${\cal
M}^{gg}_{box}$ are the amplitudes of self-energy diagrams
Figs.1(a)--(e), vertex diagrams Figs.1(f), (g) and (h) and box
diagrams Figs.1(i)--(o), respectively. They can be further
expressed as
\begin{eqnarray}
{\cal M}^{gg}_{self}=\sum_{i=1}^{14} f^i_{s} {\cal M}^{gg}_i\ ,
\hspace{1.0truecm}
\end{eqnarray}
\begin{eqnarray}
{\cal M}^{gg}_{vertex}=\sum_{i=1}^{30} f^i_{v}{\cal M}^{gg}_i\ ,
\hspace{1.0truecm}
\end{eqnarray} and
\begin{eqnarray}
{\cal M}^{gg}_{box}=\sum_{i=1}^{40} f^i_{b} {\cal M}^{gg}_i\ ,
\hspace{1.0truecm}
\end{eqnarray} respectively.
Here $f^i_{s}$, $f^i_{v}$ and $f^i_{b}$ are form factors
corresponding to the self-energy diagrams, vertex diagrams and box
diagrams, respectively, and their expressions are given explicitly
in Appendix B. ${\cal M}^{gg}_i$ are the standard matrix elements,
which are defined by
\begin{eqnarray}
&&{\cal M}^{gg}_{1,2}= \bar u(p_t)\not{\varepsilon}(k_1)k_1\cdot
\varepsilon(k_2)P_{L,R}v(p_c),\nonumber \\
&&{\cal M}^{gg}_{3,4}= \bar u(p_t)\not{\varepsilon}(k_2)k_2\cdot
\varepsilon(k_1)P_{L,R}v(p_c),\nonumber \\&&{\cal M}^{gg}_{5,6}=
\bar u(p_t)\not{\varepsilon}(k_2)p_t\cdot
\varepsilon(k_1)P_{L,R}v(p_c),\nonumber \\&&{\cal M}^{gg}_{7,8}=
\bar u(p_t)\not k_1{\varepsilon}(k_1)\cdot
\varepsilon(k_2)P_{L,R}v(p_c),\nonumber\\ &&{\cal M}^{gg}_{9,10}=
\bar u(p_t){\varepsilon}(k_1)\cdot
\varepsilon(k_2)P_{L,R}v(p_c),\nonumber \\ &&{\cal M}^{gg}_{11,12}=
\bar u(p_t)\not{\varepsilon}(k_1)\not k_1\not
\varepsilon(k_2)P_{L,R}v(p_c),\nonumber \\ &&{\cal M}^{gg}_{13,14}=
\bar u(p_t)\not{\varepsilon}(k_1)\not
\varepsilon(k_2)P_{L,R}v(p_c),\nonumber \\ &&{\cal M}^{gg}_{15,16}=
\bar u(p_t)\not{\varepsilon}(k_1)\not k_1 k_1 \cdot
\varepsilon(k_2)P_{L,R}v(p_c),\nonumber \\
&&{\cal M}^{gg}_{17,18}= \bar u(p_t)\not{\varepsilon}(k_1)p_t\cdot
\varepsilon(k_2)P_{L,R}v(p_c),\nonumber\\
&&{\cal M}^{gg}_{19,20}= \bar u(p_t)\not{\varepsilon}(k_1)\not k_1
p_t\cdot
\varepsilon(k_2)P_{L,R}v(p_c),\nonumber\\
&&{\cal M}^{gg}_{21,22}= \bar u(p_t)\not{\varepsilon}(k_2)\not k_1
k_2\cdot
\varepsilon(k_1)P_{L,R}v(p_c), \nonumber \\
&&{\cal M}^{gg}_{23,24}= \bar u(p_t)\not{\varepsilon}(k_2)\not k_1
p_t\cdot
\varepsilon(k_1)P_{L,R}v(p_c),\nonumber\\
&&{\cal M}^{gg}_{25,26}= \bar
u(p_t)k_1\cdot{\varepsilon}(k_2)p_t\cdot
\varepsilon(k_1)P_{L,R}v(p_c),\nonumber\\
&&{\cal M}^{gg}_{27,28}= \bar
u(p_t)p_t\cdot{\varepsilon}(k_2)k_2\cdot
\varepsilon(k_1)P_{L,R}v(p_c),\nonumber\\
&&{\cal M}^{gg}_{29,30}= \bar
u(p_t)p_t\cdot{\varepsilon}(k_2)p_t\cdot
\varepsilon(k_1)P_{L,R}v(p_c),\nonumber\\
&&{\cal M}^{gg}_{31,32}= \bar u(p_t)\not k_1
k_1\cdot{\varepsilon}(k_2)k_2\cdot
\varepsilon(k_1)P_{L,R}v(p_c),\nonumber\\
&&{\cal M}^{gg}_{33,34}= \bar u(p_t)\not k_1 k_1\cdot
{\varepsilon}(k_2)p_t\cdot
\varepsilon(k_1)P_{L,R}v(p_c),\nonumber\\
&&{\cal M}^{gg}_{35,36}= \bar
u(p_t)k_1\cdot{\varepsilon}(k_2)k_2\cdot
\varepsilon(k_1)P_{L,R}v(p_c),\nonumber\\
&&{\cal M}^{gg}_{37,38}= \bar u(p_t)\not k_1 p_t\cdot
{\varepsilon}(k_2)k_2\cdot
\varepsilon(k_1)P_{L,R}v(p_c),\nonumber\\
&&{\cal M}^{gg}_{39,40}= \bar u(p_t)\not k_1 p_t\cdot
{\varepsilon}(k_2)p_t\cdot \varepsilon(k_1)P_{L,R}v(p_c),
\end{eqnarray}
where $k_{1,2}$ denote the momenta of incoming partons, while
$p_{t}$ and $p_c$ are used for the outgoing top and anti-charm
quarks, respectively.

For the quarks initiated subprocesses, we also define the standard
matrix elements as
\begin{eqnarray}
&&{\cal M}^{q'\bar{q}}_{1\alpha\beta}= \bar v(k_{2})P_\alpha\gamma_{\mu}
u(k_1)\bar u(p_{t})P_\beta\gamma^{\mu} v(p_c), \nonumber \\
 &&{\cal M}^{q'\bar{q}}_{2\alpha\beta}= \bar
u(p_{t})P_\alpha v(p_c)\bar v(k_{2})P_\beta{\not p_t}
u(k_1), \nonumber \\
&&{\cal M}^{q'\bar{q}}_{3\alpha\beta}= \bar v(k_{2})P_\alpha u(k_1)\bar
u(p_{t})P_\beta
v(p_c), \nonumber \\
&&{\cal M}^{q'\bar{q}}_{4\alpha\beta}= \bar v(k_{2})P_\alpha u(k_1)\bar
u(p_{t})P_\beta {\not k_1}v(p_c),\nonumber \\
&&{\cal M}^{q'\bar{q}}_{5\alpha\beta}= \bar v(k_{2})P_\alpha v(p_c)\bar
u(p_{t})P_\beta
u(k_1), \nonumber \\
&&{\cal M}^{q'\bar{q}}_{6\alpha\beta}= \bar v(k_{2})P_\alpha v(p_c)\bar
u(p_{t})P_\beta {\not k_2}u(k_1), \nonumber \\
&&{\cal M}^{q'\bar{q}}_{7\alpha\beta}= \bar u(p_{t})P_\alpha u(k_1)\bar
v(k_{2})P_\beta{\not k_1} v(p_c),\nonumber\\
&&{\cal M}^{q'\bar{q}}_{8\alpha\beta}= \bar v(k_{2})P_\alpha{\not k_1}
v(p_c)\bar u(p_{t})P_\beta {\not
k_2}u(k_1),\nonumber\\
&&{\cal M}^{q'\bar{q}}_{9\alpha\beta}= \bar v(k_{2})P_\alpha{\not
p_t}u(k_1)\bar
u(p_{t})P_\beta{\not k_1} v(p_c),\nonumber\\
&&{\cal M}^{q'\bar{q}}_{10\alpha\beta}= \bar
v(k_{2})P_\alpha{\gamma_{\mu}} v(p_c)\bar
u(p_{t})P_\beta{\gamma^{\mu}} u(k_1),\nonumber\\
&&{\cal M}^{q'\bar{q}}_{11\alpha\beta}= \bar u(p_{t})P_\alpha
u(k_2) \bar v(p_{c})P_\beta
v(k_1),\nonumber\\
&&{\cal M}^{q'\bar{q}}_{12\alpha\beta}= \bar u(p_{t})P_\alpha
u(k_2)
\bar v(p_{c})P_\beta{\not k_2} v(k_1),\nonumber\\
&&{\cal M}^{q'\bar{q}}_{13\alpha\beta}= \bar u(p_{t})P_\beta{\not k_1}
u(k_2) \bar v(p_{c})P_\alpha v(k_1),\nonumber\\
&&{\cal M}^{q'\bar{q}}_{14\alpha\beta}= \bar u(p_{t})P_\alpha{\not
k_1} u(k_2) \bar v(p_{c})P_\beta{\not
k_2} v(k_1),\nonumber\\
&&{\cal M}^{q'\bar{q}}_{15\alpha\beta}= \bar u(p_{t})P_\alpha
{\gamma_{\mu}}u(k_2) \bar v(p_{c})P_\beta{\gamma^{\mu}} v(k_1),
\end{eqnarray} where $(\alpha$,$\beta)$=(L,L), (L,R),
(R,L) and (R,R), respectively. And the amplitude for the quarks
initiated subprocesses can be expressed as
\begin{eqnarray}
&& \hspace{-0.5cm}{\cal M}^{q'\bar{q}}=
\sum_{i=1}^{15}\sum_{\alpha,\beta=L,R}g^{i\alpha\beta}{\cal
M}^{q'\bar{q}}_{i\alpha\beta},\end{eqnarray} where the
$g^{i\alpha\beta}$ are form factors which can be fixed by the
straightforward calculations of the relevant diagrams. However,
the Feynman diagrams contributing to the subprocesses with the
different initial states can be different. We first describe the
amplitude of the subprocess $c\bar{c}\rightarrow t\bar{c}$, which
has the largest set of Feynman diagrams, i.e. all diagrams in
Fig.2, and the explicit expressions of the non-zero
$g^{i\alpha\beta}$ of each diagram can be found in Appendix B. In
Table 1 we list all other channels that can contribute to $t
\bar{c}(\bar{u})$ production and their related Feynman diagrams,
and their corresponding form factors can be obtained from ones of
$c\bar{c}\rightarrow t\bar{c}$ by modifying some couplings as
shown in the Table.

\begin{table}[ht]
\begin{center}
\begin{tabular}{ccc}
\hline\hline
channel & related diagrams in Fig.2 & form factors\\
\hline $u\bar{u} \rightarrow t\bar{u}$ & {$b_1,b_2,b_3,b_4,
v_1,v_2,v_3,v_4,s_1,s_2,s_3,s_4$}
 & $g_\rho^{i\alpha\beta}(V_{5L(R)} \leftrightarrow V_{4L(R)})$ \\
$u\bar{u} \rightarrow t\bar{c}$ & & $g_\rho^{i\alpha\beta}(V_{5L(R)} \leftrightarrow V_{4L(R)})$ \\
$c\bar{c} \rightarrow t\bar{u}$ & \raisebox{0.4cm}[0pt]{$b_1,b_2,b_3,b_4, v_1,v_2,s_1,s_2$}
& $g_\rho^{i\alpha\beta}(V_{5L(R)} \leftrightarrow V_{4L(R)})$ \\
$u\bar{c} \rightarrow t\bar{c}$ & & $g_\rho^{i\alpha\beta}(V_{5L(R)} \leftrightarrow V_{4L(R)})$\\
$c\bar{u} \rightarrow t\bar{u}$ & \raisebox{0.4cm}[0pt]{$b_1,b_2,b_3,b_4, v_3,v_4,s_3,s_4$}
 & $g_\rho^{i\alpha\beta}(V_{5L(R)} \leftrightarrow V_{4L(R)})$ \\
$c\bar{u} \rightarrow t\bar{c}$ & & $g_\rho^{i\alpha\beta}(V_{5L(R)} \leftrightarrow V_{4L(R)})$ \\
$u\bar{c} \rightarrow t\bar{u}$ & \raisebox{0.4cm}[0pt]{$b_1,b_2,b_3,b_4$}
 &$g_\rho^{i\alpha\beta}(V_{5L(R)} \leftrightarrow V_{4L(R)})$ \\
$s\bar{s} \rightarrow t\bar{c}(\bar{u})$ & & $g_\rho^{i\alpha\beta}(V_{6L(R)} \leftrightarrow V_{5L(R)})$ \\
$d\bar{d} \rightarrow t\bar{c}(\bar{u})$ & \raisebox{0.4cm}[0pt]{$b_2,b_3,v_1,v_2,s_1,s_2$}
 & $g_\rho^{i\alpha\beta}(V_{6L(R)} \leftrightarrow V_{4L(R)})$ \\ \hline\hline
\end{tabular}
\caption{ The form factors of other channels obtained from ones of
$c\bar{c} \rightarrow t\bar{c}$. Here, $V_{iL(R)}$ are defined in
the Appendix \ref{sec:cp}. The subscript $\rho$ represents the
corresponding diagram.}
\end{center}
\end{table}

The partonic level cross section for $\kappa\lambda \rightarrow
t\bar{c}$ is
\begin{equation}
\hat{\sigma}^{\kappa\lambda}
=\int_{\hat{t}_{-}}^{\hat{t}_{+}}\frac{1}{16\pi \hat{s}^2}
\overline{\sum}|{\cal M}^{\kappa\lambda}|^{2} d\hat{t}
\end{equation}
with
\begin{eqnarray}
\hat{t}_{\pm} &=& \frac{m_{c}^{2} +m_{t}^{2} -\hat{s}}{2} \pm
\frac{1}{2}\sqrt{(\hat{s} -(m_{c} +m_{t})^{2})(\hat{s} -(m_{c}
-m_{t})^{2})},
\end{eqnarray}
where $\sqrt{\hat{s}}$ is the c.m. energy of the  $\kappa \lambda$
($gg$ or $q'\bar{q}$) states. The total hadronic cross section for
$pp \rightarrow \kappa \lambda \rightarrow t\bar{c}$ can be
written in the form
\begin{equation}
\sigma(s) =\sum_{\kappa,\lambda} \int_{(m_{c}
+m_{t})/\sqrt{s}}^{1}dz \frac{dL}{dz}
\hat{\sigma}^{\kappa\lambda}(\kappa\lambda \rightarrow t\bar c \ \
{\rm at} \ \ \hat{s} =z^{2}s).
\end{equation}
Here $\sqrt{s}$ is the c.m. energy of the $pp$ states, and $dL/dz$
is the parton luminosity, defined as
\begin{equation}
\frac{dL}{dz} =2z\int_{z^{2}}^{1}
\frac{dx}{x}f_{\kappa/p}(x,\mu)f_{\lambda/p} (z^{2}/x,\mu),
\end{equation}
where $f_{\kappa/p}(x,\mu)$ and $f_{\lambda/p}(z^{2}/x,\mu)$ are
the $\kappa$ and $\lambda$ parton distribution functions,
respectively.

\section{\label{sec:dis}Numerical calculation and discussion}In the following
we present some numerical results for the total cross section of
the single top quark production induced by SUSY-QCD FCNC couplings
at the LHC. In our numerical calculations the SM parameters were
taken to be $m_t=174.3$ GeV, $M_W=80.423$ GeV, $M_Z=91.1876$ GeV,
$m_c=1.2$ GeV, $\sin^2(\theta_W)=0.23113$ and
$\alpha_s(M_Z)=0.1172$ \cite{pdg}. And we used the CTEQ6L PDF
\cite{pdf} and took the factorization scale and the
renormalization scale as $\mu_f=\mu_r=m_t/2$. The relevant SUSY
parameters are $\mu$, $\tan\beta$, $M_{\rm SUSY}$ and
$m_{\tilde{g}}$, which are unrelated to flavor changing mechanism,
and may be fixed from flavor conserving observables at the future
colliders. And they are chosen as follows: $M_{\rm{SUSY}}=400,
1000$ GeV, $\tan \beta =3, 30$, $m_{\tilde{g}}=200, 300$ GeV and
$\mu = 200$ GeV. As for the range of the flavor mixing parameters,
$(\delta^{ij}_{U})_{LL}$ are constrained by corresponding
$(\delta^{ij}_{D})_{LL}$ \cite{besmer,gam,gabre,saha}, in which
$(\delta^{12}_{U})_{LL}$ also is constrained by $K$-$\bar{K}$
mixing \cite{leb}, and $D_0$-$\bar{D}_0$ mixing makes constraints
on $(\delta^{12}_{U})_{LL}$, $(\delta^{12}_{U})_{LR}$ and
$(\delta^{12}_{U})_{RL}$ \cite{wyler}. And
$(\delta^{31}_{U})_{LL}$, $(\delta^{32}_{U})_{LL}$,
$(\delta^{31}_{U})_{RL}$ and $(\delta^{32}_{U})_{RL}$ are
constrained by the chargino contributions to $B_d$-$\bar{B}_d$
mixing \cite{gabre}. Finally, there also are constraints on the up
squark mass matrix from the chargino contributions to
$b\rightarrow s\gamma$ \cite{besmer,causse}. Taking into account
above constraints, in our numerical calculations, we use the
following limits:

(i) $(\delta^{12}_{U})_{LL}$, $(\delta^{12}_{U})_{LR}$ and
$(\delta^{12}_{U})_{RL}$ are less than $0.08M_{\rm SUSY}$/(1 TeV);

(ii) $(\delta^{12}_{U})_{RR}$ and $(\delta^{13}_{U})_{LL}$ are
limited below $0.2M_{\rm SUSY}$/(1 TeV);

(iii) $(\delta^{23}_{U})_{LL}$, $(\delta^{23}_{U})_{LR}$,
$(\delta^{23}_{U})_{RL}$, $(\delta^{23}_{U})_{RR}$,
$(\delta^{13}_{U})_{LR}$, $(\delta^{13}_{U})_{RL}$ and
$(\delta^{13}_{U})_{RR}$ vary from 0 to 1.

Our results are shown in Figs.~\ref{fig:mgl}--\ref{fig:tumix}.
There are two common features of these curves: one is that the
cross sections increase rapidly with the mixing parameters
increasing, and the other is that the cross sections depend
strongly on the gluino mass $m_{\tilde{g}}$, but weakly depend on
$\tan \beta$ (so we just discuss the results for $\tan \beta=30$
below). Comparing with above cases, the dependence on $M_{\rm
SUSY}$ is medium.

Fig.~\ref{fig:mgl} shows that in general the production rates of
gluon fusion processes are several times larger than ones of the
$q\bar{q'}$ annihilation, and these rates all depend strongly on
the gluino mass $m_{\tilde{g}}$. For example, assuming
$(\delta_{U}^{13})_{LR}=0.7$ and $\tan \beta=30$, the production
rates of $gg\rightarrow t\bar{u}$ decrease from 700 fb at
$m_{\tilde{g}}=200$ GeV to 70 fb at $m_{\tilde{g}}=500$ GeV.

Fig.~4 shows the dependence of the cross sections on the
$(\delta_{U}^{23})$ in the LL block. For $\tan \beta=30$ the cross
section of $t \bar{c}$ production can reach around 12 fb when
$(\delta_{U}^{23})_{LL} = 0.7$, $M_{\rm SUSY}=400$ GeV, and
$m_{\tilde{g}}=200$ GeV, as shown in Fig.4(b), but for $t \bar{u}$
production it is one order of magnitude smaller. The production
cross sections arising from $ (\delta_{U}^{13})_{LL}$ are less
than 1 fb as $ (\delta_{U}^{13})_{LL}$ is limited below $0.2M_{\rm
SUSY}$/(1 TeV), so we do not show the corresponding curves here.

Figs.~5 and 6 give the cross sections as the functions of
$(\delta_{U}^{13})$ and $(\delta_{U}^{23})$ in the RR block,
respectively. We can see that the cross sections of $t\bar{u}$
production are larger than ones of $t\bar{c}$ production for the
case where $(\delta_{U}^{13})_{RR} \neq 0$ and all the other
mixing parameters are set to be 0, and in contrast with this case,
the cross sections of $t\bar{c}$ production are larger than ones
of $t\bar{u}$ production for the case of $(\delta_{U}^{23})_{RR}
\neq 0$. For example, assuming $\tan \beta=30$, $M_{\rm SUSY}=400$
GeV, and $m_{\tilde{g}}=200$ GeV, the cross sections of $t
\bar{c}$ and $t \bar{u}$ productions are 12 fb and 1fb for $
(\delta_{U}^{23})_{RR} = 0.7$, respectively, but 2 fb and 13 fb
for $(\delta_{U}^{13})_{RR} = 0.7$, respectively.

Figs.~7 and 8 show the dependence of the cross sections on
$(\delta_{U}^{13})$ and $(\delta_{U}^{23})$ in the LR block,
respectively. We can see that the cross sections are much larger
than ones arising from the mixing in the LL and RR blocks, which
indicates that the SUSY FCNC effects induced by the mixing in the
LR block can enhance the cross sections significantly. For
example, for $\tan \beta=30$, $M_{\rm SUSY}=400$ GeV, and
$m_{\tilde{g}}=200$ GeV, the cross section can reach 643 fb for $
(\delta_{U}^{23})_{LR} = 0.7$, and 75 fb for $
(\delta_{U}^{13})_{LR} = 0.7$, as shown in Fig.7(b) and 8(b),
respectively. Here we see again that the production cross sections
of $t \bar{c}$ and $t \bar{u}$ depend strongly on $
(\delta_{U}^{23})_{LR} $ and $ (\delta_{U}^{13})_{LR} $,
respectively.  The cases of the RL block are similar to ones of
the LR block, so we do not show them here. From Figs.~7 and 8 we
can find that the production rates may be as large as a few pb for
favorable parameter values.

In the above discussion for Figs. 3-8, we only show the results
for $m_{\tilde{g}}=200$ GeV and $M_{\rm SUSY}=400$ GeV, and the
corresponding results for other cases ($m_{\tilde{g}}=200$ GeV and
$M_{\rm SUSY}=1000$ GeV, $m_{\tilde{g}}=300$ GeV and $M_{\rm
SUSY}=400$ or 1000 GeV) decrease significantly, but they are still
remarkable and vary from a few fb to hundreds of fb.

In Figs.~9, we give the results to uncover the interference
effects between different entries within one block and between
different blocks for $t \bar{c}$ production. Figs.~9(a)-(b) show
the typical interference effects between $ (\delta_{U}^{13})_{LR}
$ and $(\delta_{U}^{23})_{LR} $ within the LR block and between $
(\delta_{U}^{13})_{RR} $ and $(\delta_{U}^{23})_{RR} $ within the
RR block, respectively. Figs.~9(c)-(d) show the interference
effects between the LL and LR block and between the LL and RR
block for the typical parameters, respectively. In general, those
interference effects enhance the cross sections. And the
interference effects for $t \bar{u}$ production are very similar
to the case of $t \bar{c}$ production and are shown in Fig.~10.
Although the interference effects do not change the cross sections
of single top quark production at the LHC significantly, the
consideration of them is of importance to reveal the details of
flavor changing mechanism.

Finally, for convenience, we list some typical results in Table 2.
\begin{table}[ht]
\begin{center}
\begin{tabular}{cccccc}
\hline\hline subprocess & $(\delta^{23}_{U})_{LL}$ &
$(\delta^{13}_{U})_{RR}$ & $(\delta^{23}_{U})_{RR}$ &
$(\delta^{13}_{U})_{LR}$ & $(\delta^{23}_{U})_{LR}$ \\
\hline
$gg \rightarrow t\bar{u}$ & 0 & 14.4 & 0 & 650.4 & 0 \\
$q\bar{q'} \rightarrow t\bar{u}$ & 0.2 & 4.7 & 0.2 & 124.1 & 12.6 \\
$gg \rightarrow t\bar{c}$ & 11.0 & 0 & 10.5 & 0 & 633.0 \\
$q\bar{q'} \rightarrow t\bar{c}$ & 1.6 & 1.8 & 1.5 & 74.5 & 10.0 \\
\hline total & 12.8 & 20.9 & 12.2 & 849.0 & 655.6 \\
\hline\hline
\end{tabular}
\caption{Cross sections in fb for the $t\bar{c}$ and $t\bar{u}$
productions. Here $\tan \beta=30$, $M_{\rm SUSY}=400$ GeV and
$m_{\tilde{g}}=200$ GeV, the off-diagonal element in the table
equals to 0.7 and others are set to zero. }
\end{center}
\end{table}

\section{\label{sec:con}Conclusions}

We have evaluated the $t\bar{c}$ and $t\bar{u}$ productions at the
LHC within the general unconstrained MSSM framework. These single
top quark productions are induced by SUSY-QCD FCNC couplings and
have remarkable cross sections for favorable parameter values
allowed by current low energy data, which can be as large as a few
pb. It has been pointed out \cite{young1} that the top quark FCNC
production signals are more accessible than the top quark FCNC
decay signals. And according to the model independent analysis in
Ref.~\cite{young2}, it is possible to observe FCNC effects through
single top quark productions at the LHC. Thus, our above results
show that once large rates of the $t\bar{c}$ and $t\bar{u}$
productions are detected at the LHC, they may be induced by SUSY
FCNC couplings, and such flavor changing must come from the LR or
RL block, which is related to the soft trilinear couplings
$A_{U}$. Therefore, we believe that the precise measurement of
single top quark production cross sections at the LHC is a
powerful probe for the details of the SUSY FCNC couplings.

\vspace{2.5cm}

\hspace*{4.5cm}ACKNOWLEDGMENTS\\

This work was supported in part by the National Natural Science
Foundation of China and Specialized Research Fund for the Doctoral
Program of Higher Education.

\vspace{1cm}

\appendix

\section{\label{sec:cp}Couplings}

Here we list the relevant couplings in the amplitudes. $i, j$
stand for generation indices and $r, s$ stand for color indices.
\bea V_1&=&-ig_sT^a \ \ \ \ \ V_2=-g_sf_{abc} \ \ \  \ \ \ \
V_3=-ig_sf_{abc}\\
V_{4L}&=&-i\sqrt{2}g_sT_{rs}^aZ_{3i}^q\ \ \
V_{4R}=i\sqrt{2}g_sT_{rs}^aZ_{6i}^q\\
V_{5L}&=&-i\sqrt{2}g_sT_{rs}^aZ_{2i}^q\ \ \
V_{5R}=i\sqrt{2}g_sT_{rs}^aZ_{5i}^q\\
V_{6L}&=&-i\sqrt{2}g_sT_{rs}^aZ_{1i}^q\ \ \
V_{6R}=i\sqrt{2}g_sT_{rs}^aZ_{4i}^q\\
V_7&=&i/2g_s^2(1/3\delta_{ab}+d_{abc}T^c) \eea

\section{\label{sec:ff}Form Factors}

This appendix lists all the form factors in the amplitude of
various subprocess of $pp \rightarrow t\bar{c}$, in terms of 2,3-
and 4-points one-loop integrals \cite{loop}.  For convenience, we
define the abbreviations of one-loop integrals as following:
\begin{eqnarray*}
B^a &=& B(s, m_{\tilde{g}}^2, m_{\tilde{g}}^2) \\
B^d &=& B(t, m_{\tilde{g}}^2, m_{\tilde{g}}^2) \\
B^t &=& B(0, m_{\tilde{g}}^2, m_{\tilde{q}}^2) \\
B^q &=& B(m_t^2, m_{\tilde{g}}^2, m_{\tilde{q}}^2) \\
B^g &=& B(t, m_{\tilde{g}}^2, m_{\tilde{q}}^2) \\
B^m &=& B(u, m_{\tilde{g}}^2, m_{\tilde{q}}^2) \\
C &=& C( m_t^2, s,0,m_{\tilde{q}}^2, m_{\tilde{g}}^2, m_{\tilde{g}}^2) \\
C^{d} &=& C( m_t^2, s,0,m_{\tilde{g}}^2, m_{\tilde{q}}^2, m_{\tilde{q}}^2) \\
C^{t} &=& C( m_t^2, t,0,m_{\tilde{q}}^2, m_{\tilde{g}}^2, m_{\tilde{g}}^2) \\
C^{q} &=& C( m_t^2, t,0,m_{\tilde{g}}^2, m_{\tilde{q}}^2, m_{\tilde{q}}^2) \\
C^{g} &=& C( 0,m_t^2, s,m_{\tilde{g}}^2, m_{\tilde{q}}^2, m_{\tilde{g}}^2) \\
C^{f} &=& C( m_t^2, 0,s,m_{\tilde{g}}^2, m_{\tilde{q}}^2, m_{\tilde{g}}^2) \\
C^{m} &=& C( 0,0,t,m_{\tilde{q}}^2, m_{\tilde{q}}^2, m_{\tilde{g}}^2) \\
C^{n} &=& C(0,0,t,m_{\tilde{g}}^2, m_{\tilde{g}}^2, m_{\tilde{q}}^2) \\
C^{l} &=& C( 0,t,0,m_{\tilde{g}}^2, m_{\tilde{g}}^2, m_{\tilde{q}}^2) \\
C^{o} &=& C( 0,t,0,m_{\tilde{g}}^2, m_{\tilde{q}}^2, m_{\tilde{g}}^2) \\
C^{p} &=& C(0,u,0,m_{\tilde{g}}^2, m_{\tilde{q}}^2, m_{\tilde{g}}^2) \\
C^{r} &=& C( 0,u,0,m_{\tilde{q}}^2, m_{\tilde{g}}^2, m_{\tilde{q}}^2) \\
C^{x} &=& C( 0,u,m_t^2,m_{\tilde{g}}^2, m_{\tilde{g}}^2, m_{\tilde{q}}^2) \\
C^{y} &=& C(0,u, m_t^2, m_{\tilde{q}}^2, m_{\tilde{q}}^2, m_{\tilde{g}}^2) \\
C^{z} &=& C( m_t^2, t,0,m_{\tilde{g}}^2, m_{\tilde{q}}^2, m_{\tilde{g}}^2) \\
C^{w} &=& C( m_t^2, t,0,m_{\tilde{q}}^2, m_{\tilde{g}}^2,m_{\tilde{q}}^2) \\
D &=& D(0, 0, m_t^2, 0, t, s, m_{\tilde{g}}^2,m_{\tilde{q}}^2, m_{\tilde{g}}^2, m_{\tilde{q}}^2) \\
D^{v} &=& D(0, 0, m_t^2, 0, t, s, m_{\tilde{q}}^2, m_{\tilde{g}}^2, m_{\tilde{q}}^2, m_{\tilde{g}}^2) \\
D^{t} &=& D(0, m_t^2, 0, 0, u, s, m_{\tilde{q}}^2, m_{\tilde{g}}^2, m_{\tilde{q}}^2, m_{\tilde{g}}^2) \\
D^{q} &=& D( m_t^2, 0 , 0, 0, u, t, m_{\tilde{q}}^2, m_{\tilde{g}}^2, m_{\tilde{q}}^2, m_{\tilde{q}}^2) \\
D^{g} &=& D(0, 0, m_t^2, 0, t, s, m_{\tilde{g}}^2, m_{\tilde{g}}^2, m_{\tilde{q}}^2, m_{\tilde{g}}^2) \\
D^{f} &=& D(0, 0, m_t^2, 0, t, s, m_{\tilde{q}}^2, m_{\tilde{q}}^2, m_{\tilde{g}}^2, m_{\tilde{q}}^2) \\
D^{s} &=& D(0, m_t^2, 0, 0, u, s, m_{\tilde{g}}^2, m_{\tilde{g}}^2, m_{\tilde{q}}^2, m_{\tilde{g}}^2) \\
D^{m} &=& D(0, m_t^2, 0, 0, u, s, m_{\tilde{q}}^2, m_{\tilde{q}}^2, m_{\tilde{g}}^2, m_{\tilde{q}}^2) \\
D^{n} &=& D( m_t^2, 0, 0, 0, u, t, m_{\tilde{g}}^2, m_{\tilde{q}}^2, m_{\tilde{q}}^2, m_{\tilde{g}}^2) \\
D^{l} &=& D(m_t^2, 0, 0, 0, u, s, m_{\tilde{q}}^2,
m_{\tilde{g}}^2, m_{\tilde{g}}^2,m_{\tilde{q}}^2)
\end{eqnarray*}
Below we display the non-zero form factors for the gluon fusion
subprocess (we only display the odd number form factors, and the
even number ones can be obtained by the replacement of $V_{4L}
\leftrightarrow V_{4R }$ and $V_{5L} \leftrightarrow V_{5R }$
correspondingly). For self-energy diagram (see Figs.1(a)--(e)),
$f^{i}_s$ are:
\begin{eqnarray}
f^{1}_s&=&\frac{1}{m_t^3 \pi^2 s (m_t^2 -t) t}( V_{4L} (m_{\tilde
g} m_t^2 t B^t_0 V_1 V_{5L} (s V_1 +(t-m_t^2)V_3 )\nonumber \\
&&+m_t V_1(-B^g_0 m_{\tilde g} m_t^3 s V_1 V_{5L} + (m^2_t -t) (
m_{\tilde g} m_t B^q_0 V_{5R} \nonumber \\ &&-B^q_1 m_t^2
V_{5L})(sV_2 +t V_3 ) + m_t^2 s t V_1
V_{5R} B^g_1 ))) \\
f^{3}_s&=&\frac{1}{m_t^3 \pi^2 s (m_t^2 -t) t}(
V_{4L}(-m_t(m_t^2 -u)V_1(m_{\tilde g}m_t B^q_0 V_{5R} \nonumber \\
&&-B^q_1 m_t^2 V_{5L})(uV_3-sV_1)-m^2_t V_1 (m_{\tilde g} B^t_0 u
V_{5L} ((u-m_t^2)V_3 \nonumber \\ &&-sV_2 ) + m_t s V_1 (B^m_0
m_{\tilde g} m_t V_{5L} -
u V_{5R} B^m_1)))) \\
f^{5}_s&=&\frac{1}{8 \pi^2 }  ( V_1 V_{4L}  ( \frac{m_{\tilde
g}B^t_0 V_1 V_{5L}}{m_t t -m^3_t}+\frac{m_{\tilde g} B^t_0 V_2
V_{5L} }{m_t u - m_t^3}\nonumber \\ &&+\frac{V_1 (B^q_1 m_t^2
V_{5L} -m_{\tilde g} m_t B^q_0 V_{5R}) }{m_t^2 u}+\frac{V_2(B^q_1
m_t^2 V_{5L} -m_{\tilde g} m_t B^q_0 V_{5R})}{m_t^2 t}\nonumber \\
&&+ \frac{V_1 (t V_{5R} B^g_1 -B^g_0 m_{\tilde g}m_t
V_{5L})}{t(t-m_t^2)}+\frac{V_1 (u V_{5R} B^m_1  -B^m_0 m_{\tilde g} m_t V_{5L}) }{u(u-m_t^2)}    ) )\\
f^{7}_s&=&\frac{1}{m_t^3 \pi^2 s (m_t^2 -t) t}(
V_{4R}(-m_t(m_t^2 -u)V_1(m_{\tilde g}m_t B^q_0 V_{5L}\nonumber \\
&& -B^q_1 m_t^2 V_{5R})(uV_3-sV_1)-m^2_t V_1 (m_{\tilde g} B^t_0 u
V_{5R} ((u-m_t^2)V_3 -sV_2 ) \nonumber \\ &&+ m_t s V_1 (B^m_0
m_{\tilde g} m_t V_{5R} -
u V_{5L} B^m_1)))) \\
f^{9}_s&=& \frac{1}{16 m_t^2 \pi^2 s (m_t^2 - u)}(V_1 V_{4R} ( 2
m_{\tilde g} m_t^2 s (B^m_0 V_1 -
B^t_0 V_2)V_{5R}-(m_t^2 -u)(B^q_1 V_{5R} m^3_t \nonumber \\
&&-m_{\tilde g} m_t^2 B^q_0 V_{5L} + m_{\tilde g} m_t^2 B^t_0
V_{5R})V_3 -2 m_t^3 s V_1 V_{5L}
B^m_1)) \\
f^{11}_s&=&\frac{1}{2}f^{5}_s\\
 f^{13}_s&=&\frac{1}{16 m_t^2 \pi^2
t (t-m_t^2)(m_t^2-u)}(V_{4R} ((m_t^2 -u) V_1 V_2 (B^q_1 m_t^3
V_{5R} -m_{\tilde g}m_t^2 B^q_0
V_{5L}) \nonumber \\
&&+ m_t^2 V_1 (m_{\tilde g} (B^g_0 (m_t^2 -u)
(s+u)V_1+(m_t^2-t)t(B^m_0V_1 -B^t_0 V_2))V_{5R}\nonumber \\
&&+m_tV_1 V_{5L} (t(t-m_t^2)B^m_1 ))))
\end{eqnarray}
For vertex diagram Figs.1(f), (g) and (h), the non-zero form
factors $f^{i}_t$ are:
\begin{eqnarray}
f^{1}_v&=&\frac{1}{8 \pi^2 s t (t-m^2_t)} (V_{4L} (s t V_1 ( C^l_0
m_{\tilde{g}} m_t V_2 V_{5L} \nonumber \\ &&-( B^g_0 V_2 -2
C^l_{00}V_2 +2
C^o_{00} V_1 -(s + u)(C^l_{12} V_{2} -C^o_{12} V_1 ))V_{5R}) \nonumber \\
&&+ (m_t^2 -t)(V_{5L}(t((-C_0 m_{\tilde{g}}^2 +C_0 m^2_q +B^a_0 -2
C_{00} + C_1 m_t^2 )V_2 \nonumber \\ &&- 2 C^d_{00} V_1 )V_3 +s
V_1 ( V_2 ( B^g_0 -2 C^z_{00} + m_t^2 C^z_1 ) -2 V_1 C^w_{00}))
\nonumber \\ &&-m_{\tilde{g}} m_t V_2 V_{5R}(C_0 t V_3 + s V_1 C^z_{0})))) \\
f^{3}_v&=&\frac{1}{8 \pi^2 s u (u-m^2_t)} (V_{4L} ((m^2_t -u)
(m_{\tilde{g}}m_t V_2 V_{5R} (C_0 u V_3 -s V_1 C^x_0)\nonumber \\
&&+ V_{5L} ( u( 2 C^d_{00} V_1 - (- C_0 m_{\tilde{g}}^2 +C_0 m^2_q
+B^a_0 - 2 C_{00} +C_1 m_t^2 ) V_2 )V_3 \nonumber \\ && + s V_1
(V_2 (B^m_0 -2 C^x_{00} + m^2_t C^x_2 ) - 2 V_1 C^y_{00}))) -s u
V_1 (m_{\tilde{g}} m_t (C^p_1 V_2 + C^r_1 V_1 ) V_{5L}\nonumber \\
&& +(( B^m_0 -2 C^p_{00}
+ C^p_{12} u)V_2 +2 C^r_{00} V_1 - C^r_{12} u V_1 )V_{5R}))) \\
f^{5}_v&=&\frac{1}{8 \pi^2}  ( V_1 V_{4L} ( \frac{2 C^o_{00} V_1
V_{5R} }{t-m^2_t} + \frac{V_2 (C^l_0 m_{\tilde{g}} m_t V_{5L}
-B^g_0 V_{5R} +2 C^l_{00} V_{5R})}{m^2_t -t} \nonumber \\ && +
\frac{V_1 (- m_{\tilde{g}}
m_t V_{5L} C^r_1 -2C^r_{00} V_{5R} +C^r_{12} V_{5R})}{m^2_t -u}\nonumber \\
&& -\frac{V_2 (C^p_1 m_{\tilde{g}} m_t V_{5L} +(B^m_0-2 C^p_{00}
+C^p_{00} u) V_{5R})}{m^2_t -t} \nonumber \\ &&- \frac{1}{t} (V_1
(m_{\tilde{g}} m_t V_{5R} C^w_1 + V_{5L} ( 2 C^w_{00} -t C^w_{12}
+ m^2_t (C^w_1 + C^w_{11} + C^w_{12} ))))\nonumber \\ && + \frac{
V_2 ( V_{5L} ( B^m_0 -2 C^x_{00} +m^2_t C^x_2 ) - m_{\tilde{g}}m_t
V_{5R} C^x_0 )}{u} - \frac{2 V_1 V_{5L} C^y_{00} }{u}\nonumber \\
&& + \frac{1}{t}( V_2 (m_{\tilde{g}} m_t V_{5R} C^z_1 +V_{5L} (B^g_0 -2
C^z_{00} +t C^z_{12} -m^2_t (C^z_{11}
 + C^z_{12} ))))    ) ) \\
f^{7}_v&=&\frac{1}{8 \pi^2 s u (u-m^2_t)} (V_{4L}(s u V_1
(m_{\tilde{g}} m_t C^p_0 V_2 V_{5L} \nonumber \\ &&-( B^m_0 V_2 -2
C^p_{00}+ 2 C^r_{00} V_1  )V_{5R})+(m^2_t -u) (m_{\tilde{g}}m_t
V_2 V_{5R} (C_0 u V_3 -s V_1 C^x_0)\nonumber \\ &&+V_{5L} ( u( 2
C^d_{00} V_1 - (- C_0 m_{\tilde{g}}^2 +C_0 m^2_q +B^a_0 - 2 C_{00}
+C_1 m_t^2 ) V_2 )V_3 \nonumber \\ &&+ s V_1 (V_2 (B^m_0 -2
C^x_{00} + m^2_t C^x_2 ) - 2 V_1
C^y_{00}))))) \\
f^{9}_v&=& \frac{1}{16 \pi^2} ( V_{4R} ( \frac{1}{s(m^2_t
-u)}(2sV_1 (m_t (-B^m_0 V_2 +2 C^p_{00} V_2 \nonumber
\\ && -2 C^r_{00} V_1 )V_{5L} + m_{\tilde{g}} C^p_0 u V_2 V_{5R} )
\nonumber
\\ &&+ (m_t^2 -u) ( m_{\tilde{g}}(C_0 m^2_t V_2 +C_2 (t-u)V_2\nonumber
\\ && + (t-u) (C_1 V_2 -(C^d_0 +C^d_1 +C^d_2 )V_1 ))
V_{5L} \nonumber
\\ &&+m_t (-B^a_0 V_2 +(2 C_{00} +C_0 (m_{\tilde{g}}- m_q )
(m_{\tilde{g}}+m_q )+ (C_{11} + C_{12}) (t-u) \nonumber
\\ &&-C_1 (s+2u))V_2 +2 C^d_{00} V_1 +(C^d_1 + C^d_{11} +C^d_{12} )(t-u)
V_1)V_{5R})V_3 ) \nonumber
\\ &&+\frac{2 V_1 V_2 (m_t u V_{5R}C^x_2
-m_{\tilde{g}}u V_{5L} C^x_0)}{u})  )  \\
f^{11}_v&=& \frac{1}{16 \pi^2 }  ( V_{4L}  ( \frac{1}{(m_t^2 -u
)(m_t^2 -t)}(V_1(2(-(C^o_{00} +C^r_{00}) m_t^2 \nonumber
\\ && +C^r_{00}t +
C^o_{00}u)V_1 V_{5R} + V_2 (m_{\tilde g}m_t(C^p_0(m_t^2
-t)+C^l_0(m_t^2 -u))V_{5L} \nonumber
\\ && +(-(B^m_0 -2 C^p_{00}) (m_t^2 -t) + 2
C^l_{00} (m_t^2 -u) +B^g_0(u-m_t^2 ))V_{5R}))) +
 \nonumber
\\ && \frac{1}{tu}(V_1(V_2(V_{5L}(t(V^m_0 -2 C^x_{00} +m^2_t C^x_2 ) +u
(B^g_0 -2 C^z_{00} ) +m_t^2 uC^z_1)  \nonumber
\\ && - m_{\tilde{g}}m_t V_{5R} (t
C^x_0 + uC^z_0) ) -2 V_1 V_{5L}(u C^w_{00} + tC^y_{00})))  )
 ) \\
f^{13}_v&=&\frac{1}{16 \pi^2}  ( V_1 V_{4R}  ( \frac{2 m_t
C^r_{00} V_1 V_{5L}}{m_t^2 -u}+ m_{\tilde{g}}V_2 C^x_0 V_{5L} -
\frac{C^l_0 m_{\tilde{g}}(s+u)V_2 V_{5R}}{m_t^2 -t} \nonumber
\\ && +
\frac{1}{m^2_t-u}(V_2 (m_t(B^m_0-2C^p_{00})V_{5L} \nonumber
\\ && -m_{\tilde{g}}C^p_0uV_{5R}))+\frac{2m_t V_1
V_{5R}C^w_{00}}{t}- m_tV_2 V_{5R}C^x_2 \nonumber
\\ &&  +
\frac{1}{t}(V_2(m_{\tilde{g}}(s+u)V_{5L}C^z_0 -m_t V_{5R}(B^g_0 -2
C^z_{00} +(s+u)C^z_1))))) \\
f^{15}_v&=&-\frac{1}{8 \pi^2(m^2_t-t)} (V_1 V_{4R} (C^l_{12} m_t
V_2 V_{5L}
- m_t C^o_{12} V_1 V_{5L} \nonumber \\ &&+m_{\tilde{g}} ((C^l_0 +C^l_2 ) V_2 +C^o_2 V_1 )V_{5R} )) \\
f^{17}_v&=&-\frac{1}{8 \pi^2(m^2_t-t)} ( V_{4L} ((s
+u)V_1(C^l_{12}V_2-C^o_{12}V_1)V_{5R}
 \\ &&+ (m_t^2 -t) V_1 V_{5L} (V_2 C^x_{12} +V_1 C^y_{12})))\\
f^{19}_v&=& \frac{1}{8 \pi^2 (m^2_t -t)u} (V_{4R} (u V_1 (C^l_{12}
m_t V_2 V_{5L} -m_t C^o_{12} V_1 V_{5L}\nonumber
\\ && +m_{\tilde{g}}
(( C^l_{0} + C^l_{2} ) V_2 + C^o_2 V_1 )V_{5R}) + (m^2_t -t) V_1
(m_{\tilde{g}}V_{5L} (V_1 C^y_2 -V_2 (C^x_0 +C^x_2 ) ) \nonumber
\\ &&+ m_t
V_{5R} (V_2 (C^x_{12} +C^x_2 +C^x_{22}) +V_1 (C^y_{12} +C^y_2
+C^y_{22}))))) \\
f^{21}_v&=& \frac{1}{8 \pi^2 (m^2_t -u)} (V_1 V_{4R} (m_t
(C^p_{12} V_2 -C^r_{12} V_1 )V_{5L}\nonumber\\&&
+m_{\tilde{g}}((C^p_0 +C^p_1 )V_2 +C^r_1
V_1 )V_{5R})) \\
f^{23}_v&=& \frac{1}{8 \pi^2 t(m^2_t -u) } (V_{4R} ((m^2_t -u) V_1
(m_{\tilde{g}} V_{5L} (V_2 (C^z_0 +C^z_1 )- V_1 C^w_1 ) \nonumber
\\ &&- m_t
V_{5R} (V_1 (C^w_1 +C^w_{11} +C^w_{12})+V_2 (C^z_1 +C^z_{11}
+C^z_{12})))- t V_1(m_t (C^p_{12} V_2 \nonumber
\\ &&-C^r_{12} V_1 ) V_{5L}
+m_{\tilde{g}} ((C^p_0 +C^p_1 ) V_2 +C^r_{1}V_1 )V_{5R} ))) \\
f^{25}_v&=&   \frac{1}{4 \pi^2 s(m_t^2 - t)t} (V_{4R} (stV_1
(C^l_{12} m_t V_2 V_{5L} -m_t C^o_{12} V_1 V_{5L}  \nonumber
\\ && +m_{\tilde{g}}
((C^l_0+C^l_2)V_2 +C^o_2 V_1 )V_{5R} ) + (m_t^2 -t) (t
(m_{\tilde{g}} (( C_1+C_2 )V_2  \nonumber
\\ && -( C^d_0 +C^d_1 +C^d_2 ) V_1
)V_{5L} +m_t ((C_1 +C_{11} +C_{12} ) V_2 \nonumber
\\ &&  + (C^d_1 +C^d_{11}
+C^d_{12} ) V_1 )V_{5R} )V_3 +s V_1 (m_{\tilde{g}} V_{5L}(V_1
C^w_1 \nonumber
\\ && -V_2 (C^z_0+C^z_1 ) ) + m_t V_{5R} (V_1 (C^w_1 +C^w_{11}
+C^w_{12}) +V_2 (C^z_1 +C^z_{11} +C^z_{12} )))))) \\
f^{27}_v&=& \frac{1}{4 \pi^2 s u}(V_{4R} (s V_1 (
m_{\tilde{g}}V_{5L}(V_1 C^y_2 -V_2 (C^x_0 +C^x_2 ))  \nonumber
\\ && +m_t V_{5R}
(V_2 (C^x_{12} +C^x_2 +C^x_{22}) +V_1 (C^y_{12} +C^y_2
+C^y_{22}))) \nonumber
\\ && - u(m_{\tilde{g}}(( C_1 +C_2) V_2 -(C^d_0 +C^d_1 +
C^d_2)V_1 )V_{5L} + m_t (( C_1 +C_{11}  \nonumber
\\ && +C_{12}) V_2 +(C^d_1
+C^d_{11} +C^d_{12} ) V_1 )V_{5R} )V_3)) \\
f^{29}_v&=&\frac{1}{4 \pi^2 (m^2_t - t) u }(V_{4R} ((m^2_t -t)
V_1( m_{\tilde{g}} V_{5L} (V_2 (C^x_0 +C^x_2 ) -V_1 C^y_2 )
\nonumber
\\ && - m_t
V_{5R} (V_2 (C^x_{12} +C^x_{2} +C^x_{22}) +V_1 (C^y_{12} +C^y_2
+C^y_{22})))- u V_1 (C^l_{12} m_t V_2 V_{5L}  \nonumber
\\ && -m_t C^o_{12} V_1
V_{5L} + m_{\tilde{g}}((C^l_0 +C^l_2) V_2 +C^o_2 V_1 )V_{5R})))
\end{eqnarray}
And for box diagrams Figs.1(i)--(o), all form factors $f^{i}_b$
are non-zero:
\begin{eqnarray}
f^{1}_b&=& -\frac{1}{8 \pi^2}(V_{4L} (C^f_2 + C^g_0 - 2 D^g_{00} +
2D^g_{002} +2D^g_{003} -2 D^s_{003} +(D^s_{13} +D^s_{23})u
\nonumber \\ && + (D^g_{12} +D^g_{13} +D^g_{22} (s+u) + D^g_{23}
(s+ u)\nonumber \\ &&+ D^g_2 m_t^2)V_{5L}-D^g_0 m_{\tilde{g}} m_t V_{5R} ) V_2^2 \nonumber \\
&&+ (2D^l_{003}+C^m_2 -2 D^n_{003} +D^n_{23}u )V_1 V_{5L} V_2
+ 2(D^f_{002} + D^f_{003} -D^m_{003}) V_1^2 V_{5L}))  \\
f^{3}_b&=&-\frac{1}{8 \pi^2}(V_{4L} (((C^f_0 - C^g_0-  C^g_{1}
-C^g_{2}- 2D^g_{001} -2D^s_{00} \nonumber \\ &&+2( D^s_{001}+
D^s_{002}-( D^s_{13} + D^s_{23})u) +(-D^s_{12}\nonumber
\\ && +D^s_{13}-D^s_{22}+D^s_{23})u -D^g_{13} (s+u )
\nonumber \\ &&+D^g_{12}(t-s)) V_{5L}+ m_{\tilde{g}} m_t (-D^g_1+
D^s_1 +D^s_2 )V_{5R} ) V^2_2 \nonumber \\ &&+V_1((-2 D^l_{002}+
D^l_{2} (m_q^2 -m_{\tilde{g}}^2) + D^l_{12} m_t^2+ C^n_1
+2D^n_{002} +D^l_{22}
u)V_{5L}\nonumber \\ &&-D^l_2 m_{\tilde{g}} m_t V_{5R} )V_2+ 2(-D^f_{001} +D^m_{001} + D^m_{002})V_1^2 V_{5L}))   \\
f^{5}_b&=&-\frac{1}{8 \pi^2}  (V_{4L} (((C^f_1+C^f_2 + C^g_1+
C^g_2 +2D^g_{00} -2D^g_{002} +2 D^s_{00} \nonumber \\ &&-2
D^s_{002} +D^s_{23}u+ D^s_{22}u\nonumber \\ &&+ D^g_{23}t
 +D^g_{12}t)V_{5L}
 -  m_{\tilde{g}} m_t (D^g_2\nonumber \\ &&+ D^s_2)V_{5R} ) V^2_2
 +V_1((2 D^l_{001}+ 2D^l_{002} +(D^l_{1}+D^l_{2})(m_q^2 -m_{\tilde{g}}^2)
 -( D^l_{11}+ D^l_{12}) m_t^2\nonumber \\ &&+ C^n_0+ C^n_2 -2(D^n_{001} +D^n_{002})-(D^l_{12}+D^l_{22}) u)V_{5L}
 \nonumber \\ &&+(D^l_1+D^l_2 )m_{\tilde{g}} m_t V_{5R} )V_2- 2(D^f_{002} +D^m_{002})V_1^2 V_{5L}))   \\
f^{7}_b&=&\frac{1}{8 \pi^2}(V_{4L} ((  m_{\tilde{g}} m_t D^s_0
V_{5R} - (C^f_0 +2 D^g_{001} +2D^g_{002} +2D^g_{003} \nonumber
\\&&-2(2 D^s_{00} +D^s_{001}+D^s_{002}
+D^s_{003})+D^s_{2}u)V_{5L})V_2^2
 \nonumber \\ &&+2(-D^l_{002}-D^l_{003}+D^n_{00}+D^n_{002}+D^n_{003}V_1 V_{5L} V_2
 \nonumber \\ &&-2(D^f_{00} +D^f_{001} +D^f_{002}+D^f_{003}-D^m_{00}
-D^m_{001}-D^m_{002}-D^m_{003}) V_1^2 V_{5L})))  \\
f^{9}_b&=&\frac{1}{16 \pi^2} (V_{4R}( m_{\tilde{g}}
V_{5L}(2(C^f_0 -2 (D^g_{00}+D^s_{00})-D^s_0 u) V_2^2\nonumber \\
&& -4 ( D^l_{00}+ D^n_{00}) V_1 V_2 + 4 (D^f_{00}
+D^m_{00})V_1^2+C^d_0 V_7) \nonumber \\ &&+ m_t V_{5R} (-2 (-(
D^s_2 + D^s_3 )m_t^2 +C^f_1 +C^f_2 -2 D^g_{001}2 D^g_{002}
+2D^s_{00} \nonumber \\ &&+ D^s_{001}+m_t^2(D^s_{2}+2 D^s_3) -
D^s_2 u - D^s_3 m_t^2V^2_2-4(D^l_{00}\nonumber \\ && -D^l_{001}
-D^n_{001})V_1 V_2+ 4 (D^f_{00}
+D^f_{001}+D^f_{002}  - D^m_{001} )V^2_3- C^d_1 V_7)))) \\
f^{11}_b&=&\frac{1}{16 \pi^2} (V_2 V_{4L} (V_2 ((C^f_0 +C^g_0
-4D^g_{00}-4D^s_{00}
       +D^s_2 u +D^g_2t)V_{5L}\nonumber
\\&&-m_{\tilde{g}}m_t(D^g_0+D^s_0)V_{5R})
       -2(D^l_{00}+D^n_{00})V_{3}V_{5L}))\\
f^{13}_b&=&\frac{1}{16 \pi^2} (V_2 V_{4R} (m_t (( C^f_1 +C^f_2
+C^g_2 + 2 D^g_{00} +2D^s_{00}  - D^s_2 u )V_2 +2 D^l_{00}V_1
)V_{5R}\nonumber
\\&& -m_{\tilde{g}}(C^f_0 -C^g_0 -D^g_2 s -D^g_0 s  -( D^g_0 + D^s_0 )u) V_2 V_{5L}))\\
f^{15}_b&=&\frac{1}{8 \pi^2}  (V_2 V_{4R}(-m_{\tilde{g}} D^n_3 V_1
V_{5L} + m_t D^n_{13} V_1 V_{5R} + V_2 (  m_{\tilde{g}}(  D^g_0
\nonumber \\ &&+ D^g_2 +D^g_3 - D^s_3 ) V_{5L}- m_t ( D^g_{12} +
D^g_{13} + D^g_2
+D^g_{22} +  D^g_{23} + D^s_{13} )V_{5R})))   \\
f^{17}_b&=&\frac{1}{8 \pi^2}(((C^f_2 + C^g_2 + 2 D^g_{00} +
2D^g_{002} +2(D^s_{00} + D^s_{002}) +m^t_2 D^s_{23} - D^s_{23}
(s+t) \nonumber \\ && -(D^s_{12} +D^s_{2}+D^s_{22}+D^s_{23})u +
(D^g_{12} +D^g_{2} +D^g_{22} (s+u) \nonumber \\ && - (2 D^l_{001}
+ 2 D^l_{002} - C^m_2 -2D^n_{00} -2 D^n_{001} -2D^n_{002}\nonumber
\\ &&    + (D^n_{12}+ D^n_2 + D^n_{22})u )V_1 V_2
+ 2 (D^f_{002} + D^m_{002})V^2_3 ) V_{4L} V_{5L})   \\
f^{19}_b&=&\frac{1}{8 \pi^2}(V_2 V_{4R} (m_t((D^g_{12} + D^G_2
+D^g_{22} - D^s_{12} ) V_{2}+( D^n_{1} + D^n_{11} + D^n_{12})V_1)
V_{5R} \nonumber \\ &&-m_{\tilde{g}} ((D^g_{0} +  D^g_{2} +
D^s_{0} +D^s_2)V_{2}
+(D^n_0 +D^n_1 +D^n_2)V_1)V_{5L}))  \\
f^{21}_b&=&\frac{1}{8 \pi^2} (V_2 V_{4R}
(m_{\tilde{g}}(D^g_{1}V_2- (D^s_0 +D^s_{1} + D^s_{2} ) V_{2}+D^l_2
V_1 )V_{5L}+m_t ((D^g_{13} + \nonumber \\ &&D^s_{12} + D^s_{13}
+D^s_2+D^s_{22} +D^s_{23})V_{2} +
(D^l_{12} +D^l_2 +D^l_{22}+D^l_{23})V_1)V_{5R}))  \\
f^{23}_b&=&\frac{1}{8 \pi^2}(V_2 V_{4R} (m_{\tilde{g}}((D^g_{0} +
D^g_2 +D^s_{0}+D^s_{2} ) V_{2}\nonumber \\ &&-( D^n_{1} +
D^l_2)V_1) V_{5L} -m_t((-D^g_{23} + D^s_{2} + D^s_{22}
+D^s_{23})V_{2} \nonumber \\ && +
(D^l_{1} +D^l_{11} +2D^l_{12} +D^l_{13} +D^l_{2} +D^l_{22}+D^l_{23})V_1)V_{5R}))  \\
f^{25}_b&=&-\frac{1}{4 \pi^2}(V_{4R}( m_{\tilde{g}} ((D^g_0
-D^g_{22}-D^g_{23}+D^s_{23})V_2^2 \nonumber \\
&&+ ( D^l_{13}+ D^l_{23} + D^n_{23}+D^n_{13}) V_1 V_2 +  (D^f_{22}
+D^f_{23}-D^m_{23})V_1^2)V_{5L}\nonumber \\
&&+m_t((D^g_{122} + D^g_{123} + D^g_{22}+D^g_{222} +D^g_{223} +
D^g_{23}-D^m_{23})V^2_2\nonumber \\
&&+(D^l_{133}+D^l_{123}-D^n_{133}-D^n_{123})V_1
V_2\nonumber \\
&&+ (D^f_{122}+D^f_{123}+D^f_{22}+D^f_{222} + D^f_{223}+D^f_{23}+ D^m_{123} )V^2_3)V_{5R}))  \\
f^{27}_b&=&\frac{1}{4 \pi^2}(V_{4R}(( m_{\tilde{g}} (-D^g_{1}
-D^g_{12}+D^s_{1}+D^s_{12}+D^s_{2}+D^s_{22})V_{5L}\nonumber
\\&&+m_t(D^g_{112} +
D^g_{122} + D^g_{12}+D^s_{112} +D^s_{122})V_{5R})V^2_2+V_1(
m_{\tilde{g}}(D^l_{12}+D^l_{22}\nonumber
\\&&+D^n_{12}+D^n_{2}+D^n_{22})V_{5L}
+m_t(D^l_{112}+D^l_{12}+D^l_{122}-D^n_{112}\nonumber
\\&&-D^n_{12}-D^n_{122})V_{5R})
V_2+V_1^2((D^f_{12}-D^m_{12}-D^m_{22})m_{\tilde{g}}V_{5L}+
(D^f_{112}+D^f_{12}\nonumber
\\&& + D^f_{122}+D^m_{112} +D^m_{122})m_tV_{5R}))) \\
f^{29}_b&=&-\frac{1}{4 \pi^2} (V_{4R}(( m_{\tilde{g}} (D^g_{2}
+D^g_{22}+D^s_{2}+D^s_{22})V_{5L}\nonumber
\\&&-m_t(D^g_{122} + D^g_{22} +
D^g_{222}-D^s_{122})V_{5R})V^2_2+V_1( m_{\tilde{g}}
(D^l_{11}\nonumber
\\&&+2D^l_{12}+D^l_{22}+D^n_{1}+D^n_{11}+2D^n_{12}+D^n_{2}+D^n_{22})V_{5L}
\nonumber
\\&&+m_t(D^l_{11}+D^l_{111}+2D^l_{112}+D^l_{12}+D^l_{122}
-D^n_{11}-D^n_{111}-2D^n_{112}\nonumber
\\&&-D^n_{12}-D^n_{122})V_{5R})V_2-
V_1^2((D^f_{22} +D^m_{22})m_{\tilde{g}}V_{5L}\nonumber
\\&&+ (D^f_{122}+D^f_{22}
+ D^f_{222}+D^f_{122} -D^m_{122})m_tV_{5R})))\\
f^{31}_b&=&\frac{1}{4 \pi^2}( ((D^g_{112}+
D^g_{113}+D^g_{12}+D^g_{122}+2D^g_{123}+D^g_{13}+D^g_{133}\nonumber \\
&&-D^s_{113}-2D^s_{123}
-D^s_{13}-D^s_{133}-D^s_{23}-D^s_{223}-D^s_{233})V_2^2 \nonumber \\
&&+ (D^l_{223}+ D^l_{23} + D^l_{233}-
D^n_{23}-D^n_{223}-D^n_{233})
V_1V_2\nonumber \\
&&+(D^f_{112}+D^f_{113}+D^f_{12}+D^f_{122}+2D^f_{123}+D^f_{13}+D^f_{133}-D^m_{113}
-2D^m_{123}\nonumber \\
&&-D^m_{13}-D^m_{133}-D^m_{223}-D^m_{23}-D^m_{233})V_1^2)V_{4L}V_{5L})  \\
f^{33}_b&=&\frac{1}{4 \pi^2}( ((D^g_{12}+
D^g_{122}+D^g_{123}+D^g_{13}+D^g_{2}+2D^g_{22}+D^g_{222}+2D^g_{223}\nonumber \\
&&+2D^g_{23}+D^g_{233}
+D^s_{123}+D^s_{13}+D^s_{223}+D^s_{23}+D^s_{233})V_2^2 -
(D^l_{123}+ D^l_{13}\nonumber \\
&& + D^l_{133}+D^l_{223}+ D^l_{23} + D^l_{233}-
D^n_{123}-D^n_{223}-D^n_{133}\nonumber \\
&&-D^n_{23}-D^n_{233})V_1V_2+
(D^f_{122}+D^f_{123}+D^f_{22}+D^f_{222}+2D^f_{223}+D^f_{23}\nonumber \\
&&+D^f_{233}+D^m_{123}
+D^m_{223}+D^m_{23}+D^m_{233})V_1^2)V_{4L}V_{5L})  \\
f^{35}_b&=&\frac{1}{4 \pi^2} (V_{4R}(( m_{\tilde{g}} (D^g_{12}
+D^g_{13}+D^s_{13}+D^s_{23})V_{5L}-m_t(D^g_{112} + D^g_{113}\nonumber \\
&& + D^g_{12}+D^g_{122} +D^g_{123} +
D^g_{13}-D^s_{113}-D^s_{123})V_{5R})V^2_2\nonumber \\
&&+V_1 ((m_{\tilde{g}}
(D^l_{23}+D^n_{23})V_{5L}+m_t(D^l_{123}-D^n_{123})V_{5R}))
V_2\nonumber
\\&&-V_1^2((D^f_{12}+D^f_{13}+D^m_{13}+D^m_{23})m_{\tilde{g}}V_{5L}+
(D^f_{112}\nonumber
\\&&+D^f_{113} + D^f_{12} +D^f_{122}+D^f_{123}+D^f_{13}
-D^m_{113} -D^m_{123})m_tV_{5R})))\\
f^{37}_b&=&-\frac{1}{4 \pi^2} (((D^g_{112}+
D^g_{12}+D^g_{122}+D^g_{123}+D^g_{13}+D^s_{112}\nonumber
\\&&+2D^s_{12}
+2D^s_{122}+D^s_{123}+D^s_{13}+D^s_{2}+2D^s_{22}+D^s_{222}\nonumber
\\&&+D^s_{223}+D^s_{23})V_2^2
+(-D^l_{122}- D^l_{123} - D^l_{22}- D^l_{222} - D^l_{223}\nonumber
\\&&
+D^n_{12}+D^n_{122}+D^n_{123}+D^n_{2}+2D^n_{22}+D^n_{222}+D^n_{223}\nonumber
\\&&+D^n_{23})
V_1V_2+(D^f_{112}+D^f_{123}+D^f_{12}+D^f_{122}+D^m_{112}+D^m_{12}\nonumber
\\&&+2D^m_{122}
+D^m_{123}+D^m_{22}+D^m_{222}+D^m_{223})V_1^2)V_{4L}V_{5L}) \\
f^{39}_b&=&\frac{1}{4 \pi^2}(((-D^g_{12}-
D^g_{122}-D^g_{2}-2D^g_{22}-D^g_{222}-D^g_{223}\nonumber
\\&&-D^g_{23}+D^s_{12}+D^s_{122}
+D^s_{2}+2D^s_{22}+D^s_{222}+D^s_{223}\nonumber
\\&&+D^s_{23})V_2^2
+(-D^l_{112}- D^l_{113} - D^l_{12}- 2D^l_{122} - 2D^l_{123}
-D^l_{22}-D^l_{222}-D^l_{223}\nonumber
\\&&+D^n_{112}+D^n_{113}+2D^n_{12}+2D^n_{122}+2D^n_{123}+D^n_{13}
+D^n_{2}+2D^n_{22}\nonumber
\\&&+D^n_{222}+D^n_{223}+D^n_{23})V_1V_2+
(-D^f_{122}-D^f_{22}-D^f_{222}-D^f_{223}\nonumber
\\&&+D^m_{122}+D^m_{22}+D^m_{222}
+D^m_{223})V_1^2)V_{4L}V_{5L})
\end{eqnarray}
For the quarks initiated subprocesses, we list all form factors of
$c\bar c$ initial state:
\begin{eqnarray}
g^{6LR}_{s_4}&=&g^{14LR}_{s_4} = \frac{m_{\tilde{g}} B^t_0 V_1^2 V_{4R} V_{5R}}{8m_t \pi^2 t}  \\
g^{6RL}_{s_4}&=&g^{14LL}_{s_4} = g^{6LR}_{s_4} (V_{4L}\leftrightarrow V_{4R},V_{5L}\leftrightarrow V_{5R})\\
g^{6LR}_{s_3}&=&g^{14LR}_{s_3} = \frac{V_1^2 V_{4R} (B_1 m_t^2
V_{5R} -m_{\tilde{g}} m_tB^q_0
V_{5L})}{8m_t^2 \pi^2 t} \\
g^{6RL}_{s_3}&=&g^{14LL}_{s_3} = g^{6LR}_{s_3}
(V_{4L}\leftrightarrow V_{4R},V_{5L}\leftrightarrow
V_{5R})\\
g^{5LR}_{s_2}&=&g^{14RR}_{s_2}=\frac{V_1^2 V_{4R} (m_{\tilde{g}}
m_t B^q_0 V_{5L} -B^q_1
m_t^2V_{5R}}{8 m_t^2 \pi^2 s} \\
g^{5RL}_{s_2}&=&g^{14RL}_{s_2}=g^{5LR}_{s_2}(V_{4L}\leftrightarrow V_{4R},V_{5L}\leftrightarrow V_{5R})\\
g^{5LR}_{s_1}&=&g^{14RR}_{s_1}=-\frac{m_{\tilde{g}} B_0^t V_1^2 V_{4R} V_{5R}}{8 m_t \pi^2 s}\\
g^{5RL}_{s_1}&=&g^{14RL}_{s_1}=g^{5LR}_{s_1}(V_{4L}\leftrightarrow V_{4R},V_{5L}\leftrightarrow V_{5R})\\
g^{2LL}_{v_4}&=&g^{2LR}_{v_4}=\frac{1}{8\pi^2 t}(V_1 V_1V_{4R}(m_t(C^q_1+C^q_{11}\nonumber\\&&+C^q_{12})V_{5R}-m_{\tilde{g}}(C^q_0+C^q_1+C^q_2)V_{5L})) \\
g^{2RL}_{v_4}&=&g^{2RR}_{v_4}=g^{2LL}_{v_4}(V_{4L}\leftrightarrow V_{4R},V_{5L}\leftrightarrow V_{5R})\\
g^{6LR}_{v_4}&=&g^{14LR}_{v_4}=\frac{C^q_{00} V_1 V_1 V_{4R} V_{5R}}{4 \pi^2 t} \\
g^{6RL}_{v_4}&=&g^{14LL}_{v_4}=g^{6LR}_{v_4}(V_{4L}\leftrightarrow V_{4R},V_{5L}\leftrightarrow V_{5R})\\
g^{2LL}_{v_3}&=&g^{2LR}_{v_3}=-\frac{1}{8 \pi^2 t}(V_1 V_2 V_{4R} (m_{\tilde{g}} (C^t_1+C^t_2)V_{5L}+m_t(C^t_1+C^t_{11}+C^t_{12})V_{5R}) )\\
g^{2RL}_{v_3}&=&g^{2RR}_{v_3}=g^{2LL}_{v_3}(V_{4L}\leftrightarrow V_{4R},V_{5L}\leftrightarrow V_{5R})\\
g^{6LR}_{v_3}&=&g^{14LR}_{v_3}=-\frac{1}{8\pi^2 t}(V_1 V_2
V_{4R}(C_0^t V_{5R} m_{\tilde{g}}^2 + m_t C^t_0 V_{5L}
m_{\tilde{g}} \nonumber\\ &&-(C^t_0
m_{q}^2 +B^d_0-2 C^t_{00} +m_t^2 C^t_1 )V_{5R})) \\
g^{6RL}_{v_3}&=&g^{14LL}_{v_3}=g^{6LR}_{v_3}(V_{4L}\leftrightarrow V_{4R},V_{5L}\leftrightarrow V_{5R})\\
g^{5LR}_{v_2}&=&g^{14RR}_{v_2}=\frac{C^d_{00} V_1 V_1 V_{4R} V_{5R}}{4 \pi^2 s} \\
g^{5RL}_{v_2}&=&g^{14RL}_{v_2}=g^{5LR}_{v_2}(V_{4L}\leftrightarrow V_{4R},V_{5L}\leftrightarrow V_{5R})\\
g^{8LL}_{v_2}&=&g^{8LR}_{v_2}=\frac{1}{8 \pi^2 s}(V_1 V_1 V_{4R} ((C^d_1+C^d_{11}+C^d_{12})m_tV_{5R} \nonumber\\&&- (C^d_0+C^d_1+C^d_2)m_{\tilde{g}} V_{5L})) \\
g^{8RL}_{v_2}&=&g^{8RR}_{v_2}=g^{8LL}_{v_2}(V_{4L}\leftrightarrow V_{4R},V_{5L}\leftrightarrow V_{5R})\\
g^{5LR}_{v_1}&=&g^{14RR}_{v_1}=\frac{1}{8 \pi^2 s}(V_1 V_2 V_{4R} (C_0 m_{\tilde{g}} m_t V_{5L}\nonumber\\ && - (-C_0 m_{\tilde{g}}^2 +C_0 m^2_q+B^a_0-2C_{00}+C_1 m_2^t)V_{5R})) \\
g^{5RL}_{v_1}&=&g^{14RL}_{v_1}=g^{5LR}_{v_1}(V_{4L}\leftrightarrow V_{4R},V_{5L}\leftrightarrow V_{5R})\\
g^{8LL}_{v_1}&=&g^{8LR}_{v_1}=\frac{V_1 V_2 V_{4R} ((C_1 +C_2 )m_{\tilde{g}} V_{5L} +(C_1 +C_{11} +C_{12} )m_t V_{5R})}{8 \pi^2 s} \\
g^{8RL}_{v_1}&=&g^{8RR}_{v_1}=g^{8LL}_{v_1}(V_{4L}\leftrightarrow V_{4R},V_{5L}\leftrightarrow V_{5R})\\
g^{1LL}_{b_4}&=&\frac{m_{\tilde{g}} D^q_3 V_{4R}^3 V_{5L}}{16 \pi^2} \\
g^{1RR}_{b_4}&=&g^{1LL}_{b_4}(V_{4L}\leftrightarrow V_{4R},V_{5L}\leftrightarrow V_{5R})\\
g^{1LR}_{b_4}&=&\frac{m_{\tilde{g}} D^q_3 V_{4R}^2 V_{4L} V_{5R}}{16 \pi^2} \\
g^{1RL}_{b_4}&=&g^{1LR}_{b_4}(V_{4L}\leftrightarrow V_{4R},V_{5L}\leftrightarrow V_{5R})\\
g^{4LL}_{b_4}&=&\frac{m_{\tilde{g}} V_{4L} V_{4R}^2 (m_{\tilde{g}} D^q_0 V_{5L} +m_t (D^q_0+D^q_1+D^q_2)V_{5R})}{16 \pi^2} \\
g^{4RR}_{b_4}&=&g^{4LL}_{b_4}(V_{4L}\leftrightarrow V_{4R},V_{5L}\leftrightarrow V_{5R})\\
g^{4LR}_{b_4}&=&\frac{m_{\tilde{g}} V_{4L}^3 (m_{\tilde{g}} D^q_0 V_{5L} +m_t (D^q_0+D^q_1+D^q_2)V_{5R})}{16 \pi^2} \\
g^{4RL}_{b_4}&=&g^{4LR}_{b_4}(V_{4L}\leftrightarrow V_{4R},V_{5L}\leftrightarrow V_{5R})\\
g^{9LL}_{b_4}&=&g^{9LR}_{b_4}=\frac{D^q_{13} V_{4L} V_{4R}^2 V_{5L}}{16 \pi^2 } \\
g^{9RL}_{b_4}&=&g^{9RR}_{b_4}=g^{9LL}_{b_4}(V_{4L}\leftrightarrow V_{4R},V_{5L}\leftrightarrow V_{5R})\\
g^{11LL}_{b_4}&=&g^{11LR}_{b_4}=\frac{V_{4L}^2 V_{4R} (m_{\tilde{g}} D^q_1 V_{5L}+m_t(D^q_1+D^q_{11}+D^q_{12})V_{5R} )}{16 \pi^2} \\
g^{11RL}_{b_4}&=&g^{11RR}_{b_4}=g^{11LL}_{b_4}(V_{4L}\leftrightarrow V_{4R},V_{5L}\leftrightarrow V_{5R})\\
g^{13LL}_{b_4}&=&g^{13RL}_{b_4}=\frac{D^q_{00} V_{4L} V_{4R}^2 V_{5L} }{8 \pi^2} \\
g^{13LR}_{b_4}&=&g^{13RR}_{b_4}=g^{13LL}_{b_4}(V_{4L}\leftrightarrow V_{4R},V_{5L}\leftrightarrow V_{5R})\\
g^{1LL}_{b_3}&=&-\frac{m_{\tilde{g}} D^t_3 V_{4R}^3 V_{5L} }{16 \pi^2} \\
g^{1RR}_{b_3}&=&g^{1LL}_{b_3}(V_{4L}\leftrightarrow V_{4R},V_{5L}\leftrightarrow V_{5R})\\
g^{1LR}_{b_3}&=&-\frac{m_{\tilde{g}} D^t_3 V_{4R}^2 V_{4L} V_{5R} }{16 \pi^2} \\
g^{1RL}_{b_3}&=&g^{1LR}_{b_3}(V_{4L}\leftrightarrow V_{4R},V_{5L}\leftrightarrow V_{5R})\\
g^{4LL}_{b_3}&=&\frac{m_{\tilde{g}} V_{4L} V_{4R}^2 (m_{\tilde{g}} D^t_0 V_{5L} - m_t D^t_2 V_{5R} )}{16 \pi^2} \\
g^{4RR}_{b_3}&=&g^{4LL}_{b_3}(V_{4L}\leftrightarrow V_{4R},V_{5L}\leftrightarrow V_{5R})\\
g^{4LR}_{b_3}&=&\frac{m_{\tilde{g}} V_{4L}^3 (m_{\tilde{g}} D^t_0 V_{5L} - m_t D^t_2 V_{5R} )}{16 \pi^2} \\
g^{4RL}_{b_3}&=&g^{4LR}_{b_3}(V_{4L}\leftrightarrow V_{4R},V_{5L}\leftrightarrow V_{5R})\\
g^{9LL}_{b_3}&=&g^{9LR}_{b_3}=\frac{D^t_{13} V_{4L} V_{4R}^2 V_{5L} }{16 \pi^2} \\
g^{9RL}_{b_3}&=&g^{9RR}_{b_3}=g^{9LL}_{b_3}(V_{4L}\leftrightarrow V_{4R},V_{5L}\leftrightarrow V_{5R})\\
g^{11LL}_{b_3}&=&g^{11LR}_{b_3}=\frac{V_{4L}^2 V_{4R} (m_t D^t_{12} V_{5R} -m_{\tilde{g}} D^t_1 V_{5L})}{16 \pi^2} \\
g^{11RL}_{b_3}&=&g^{11RR}_{b_3}=g^{11LL}_{b_3}(V_{4L}\leftrightarrow V_{4R},V_{5L}\leftrightarrow V_{5R})\\
g^{13LL}_{b_3}&=&g^{13RL}_{b_3}=-\frac{D^t_{00} V_{4L} V_{4R}^2 V_{5L}}{8 \pi^2} \\
g^{13LR}_{b_3}&=&g^{13RR}_{b_3}=g^{13LL}_{b_3}(V_{4L}\leftrightarrow V_{4R},V_{5L}\leftrightarrow V_{5R})\\
g^{2LL}_{b_2}&=&\frac{V_{4L}^2 V_{4R} (m_{\tilde{g}} V_{5L} D^v_3 -m_t V_{5R} D^v_{23})}{16 \pi^2} \\
g^{2RR}_{b_2}&=&g^{2LL}_{b_2}(V_{4L}\leftrightarrow V_{4R},V_{5L}\leftrightarrow V_{5R})\\
g^{2LR}_{b_2}&=&\frac{V_{4R}^3 (m_{\tilde{g}} V_{5L} D^v_3 -m_t V_{5R} D^v_{23})}{16 \pi^2} \\
g^{2RL}_{b_2}&=&g^{2LR}_{b_2}(V_{4L}\leftrightarrow V_{4R},V_{5L}\leftrightarrow V_{5R})\\
g^{5RR}_{b_2}&=&g^{5RL}_{b_2}=\frac{m_{\tilde{g}} V_{4L}^2 V_{4R} (m_{\tilde{g}} V_{5R} D^v_0 - m_t V_{5L} D^v_2)}{16 \pi^2} \\
g^{5LL}_{b_2}&=&g^{5LR}_{b_2}=g^{5RR}_{b_2}(V_{4L}\leftrightarrow V_{4R},V_{5L}\leftrightarrow V_{5R})\\
g^{6LR}_{b_2}&=&\frac{V_{4L}^2 V_{4R} V_{5R} D^v_{00}}{8 \pi^2} \\
g^{6RL}_{b_2}&=&g^{6LR}_{b_2}(V_{4L}\leftrightarrow V_{4R},V_{5L}\leftrightarrow V_{5R})\\
g^{10LL}_{b_2}&=&-\frac{V_{4L}^3 V_{5L} D^v_{13}}{16 \pi^2} \\
g^{10RR}_{b_2}&=&g^{10LL}_{b_2}(V_{4L}\leftrightarrow V_{4R},V_{5L}\leftrightarrow V_{5R})\\
g^{10LR}_{b_2}&=&-\frac{V_{4L}^2 V_{4R} V_{5L} D^v_{13}}{16 \pi^2} \\
g^{10RL}_{b_2}&=&g^{10LR}_{b_2}(V_{4L}\leftrightarrow V_{4R},V_{5L}\leftrightarrow V_{5R})\\
g^{12LL}_{b_2}&=&g^{12LR}_{b_2}=-\frac{m_{\tilde{g}} V_{4L}^2 V_{4R} V_{5L} D^v_1}{16 \pi^2} \\
g^{12RL}_{b_2}&=&g^{12RR}_{b_2}=g^{12LL}_{b_2}(V_{4L}\leftrightarrow V_{4R},V_{5L}\leftrightarrow V_{5R})\\
g^{14LL}_{b_2}&=&\frac{V_{4L}^3 V_{5L} D^v_{00}}{8 \pi^2} \\
g^{14LR}_{b_2}&=&g^{14LL}_{b_2}(V_{4L}\leftrightarrow V_{4R},V_{5L}\leftrightarrow V_{5R})\\
g^{3LL}_{b_1}&=&g^{3RL}_{b_1}=\frac{(D_0+D_1 +D_2 +D_3)m_{\tilde{g}} V_{4L}^2 V_{4R} V_{5L}}{16 \pi^2} \\
g^{3LR}_{b_1}&=&g^{3RR}_{b_1}=g^{3LL}_{b_1}(V_{4L}\leftrightarrow V_{4R},V_{5L}\leftrightarrow V_{5R})\\
g^{5LR}_{b_1}&=&-\frac{D_{00} V_{4L}^2 V_{4R} V_{5R} }{8 \pi^2} \\
g^{5RL}_{b_1}&=&g^{5LR}_{b_1}(V_{4L}\leftrightarrow V_{4R},V_{5L}\leftrightarrow V_{5R})\\
g^{6LL}_{b_1}&=&g^{6RL}_{b_1}=-\frac{m_{\tilde{g}} V_{4L} V_{4R}^2 (D_0 m_{\tilde{g}}V_{5L} +(D_0 + D_1 +D_2)m_t V_{5R})}{16 \pi^2} \\
g^{6LR}_{b_1}&=&g^{6RR}_{b_1}=g^{6LL}_{b_1}(V_{4L}\leftrightarrow V_{4R},V_{5L}\leftrightarrow V_{5R})\\
g^{7LL}_{b_1}&=&\frac{(D_{12}+D_2+D_{22} +D_{23})V_{4L}^3 V_{5L}}{16 \pi^2} \\
g^{7RR}_{b_1}&=&g^{7LL}_{b_1}(V_{4L}\leftrightarrow V_{4R},V_{5L}\leftrightarrow V_{5R})\\
g^{7LR}_{b_1}&=&\frac{(D_{12}+D_2+D_{22} +D_{23})V_{4L}^2 V_{4R} V_{5R}}{16 \pi^2} \\
g^{7RL}_{b_1}&=&g^{7LR}_{b_1}(V_{4L}\leftrightarrow V_{4R},V_{5L}\leftrightarrow V_{5R})\\
g^{8LL}_{b_1}&=&-\frac{V_{4L}^2 V_{4R} (D_2 m_{\tilde{g}} V_{5L} +(D_{12} +D_2 +D_{22} )m_t V_{5R} )}{16 \pi^2} \\
g^{8RR}_{b_1}&=&g^{8LL}_{b_1}(V_{4L}\leftrightarrow V_{4R},V_{5L}\leftrightarrow V_{5R})\\
g^{8LR}_{b_1}&=&-\frac{V_{4R}^3 (D_2 m_{\tilde{g}} V_{5L} +(D_{12} +D_2 +D_{22} )m_t V_{5R} )}{16 \pi^2} \\
g^{8RL}_{b_1}&=&g^{8LR}_{b_1}(V_{4L}\leftrightarrow V_{4R},V_{5L}\leftrightarrow V_{5R})\\
g^{14RL}_{b_1}&=&-\frac{D_{00} V_{4L}^3 V_{5L} }{8 \pi^2} \\
g^{14RR}_{b_1}&=&g^{14RL}_{b_1}(V_{4L}\leftrightarrow V_{4R},V_{5L}\leftrightarrow V_{5R})\\
\end{eqnarray}

\newpage
\begin{figure}
\centerline{\epsfig{file=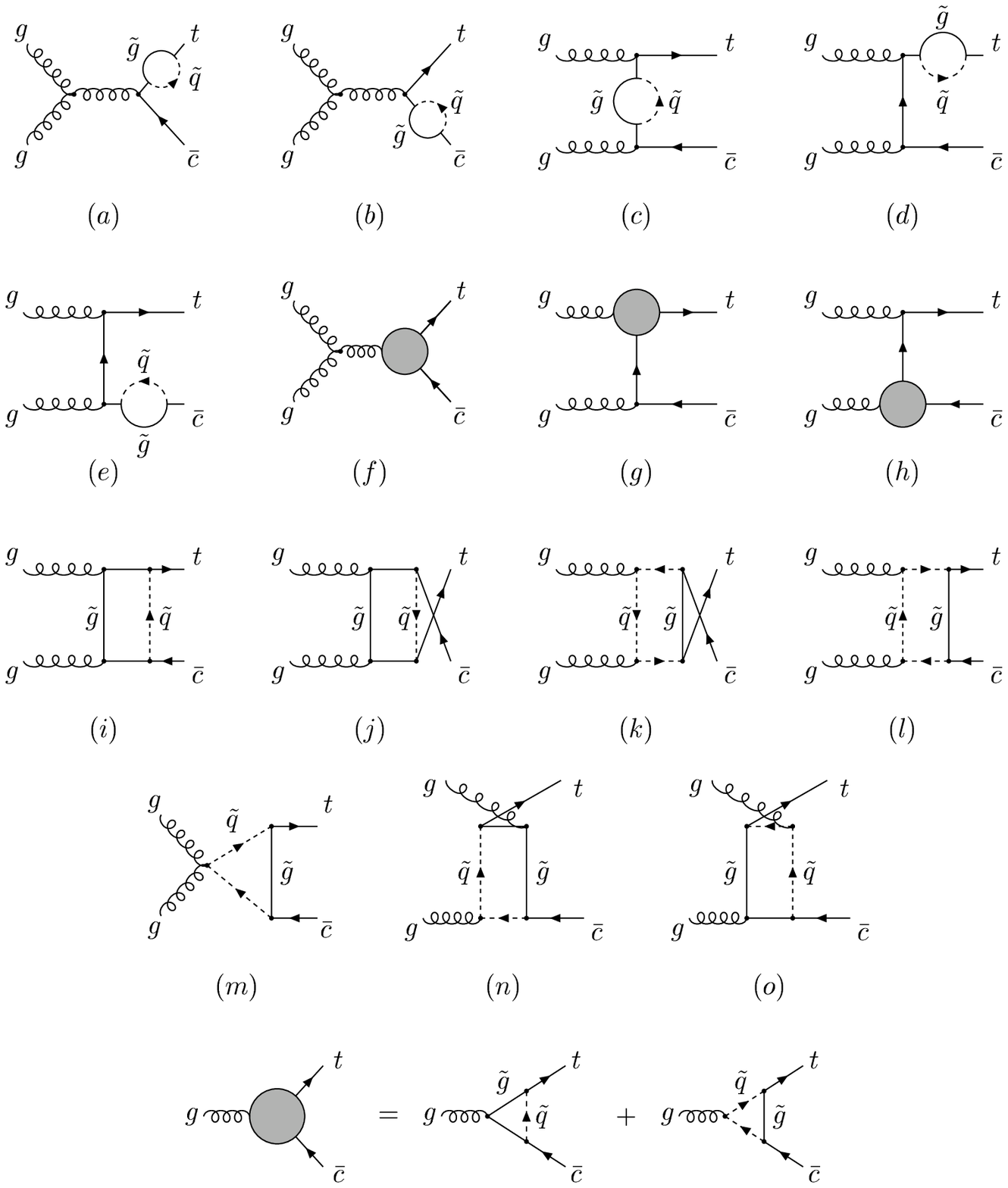,width=400pt}}
\caption[]{Gluon initial state subprocess Feynman diagrams for the
top quark and anti-charm quark associated production.
\label{fig:feyn}}
\end{figure}

\newpage
\begin{figure}
\centerline{\epsfig{file=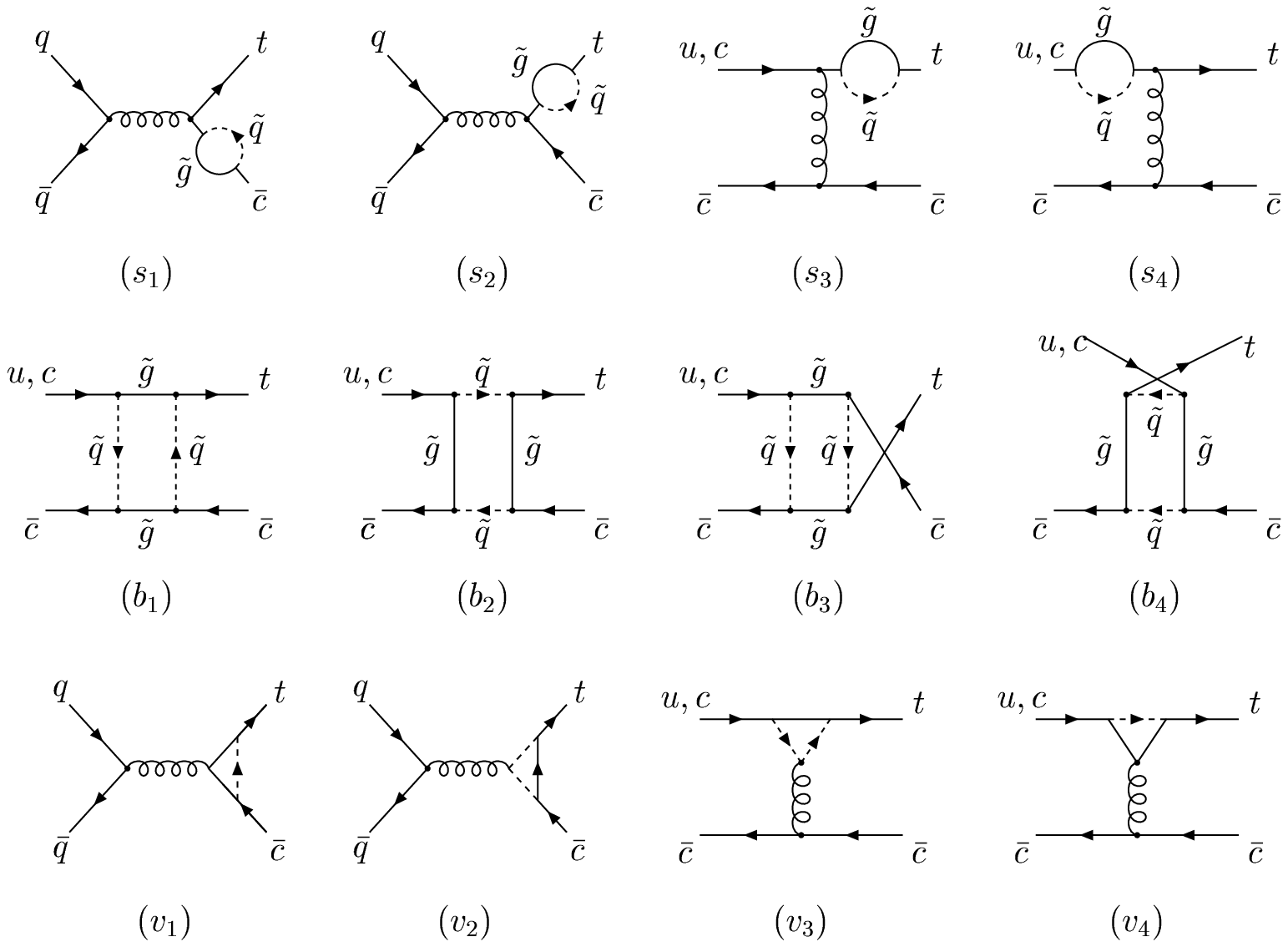,width=400pt}}
\caption[]{Quark initial state subprocess Feynman diagrams for the
top quark and anti-charm quark associated production.
\label{fig:feynmanq}}
\end{figure}
\newpage

\begin{figure}
\centerline{\epsfig{file=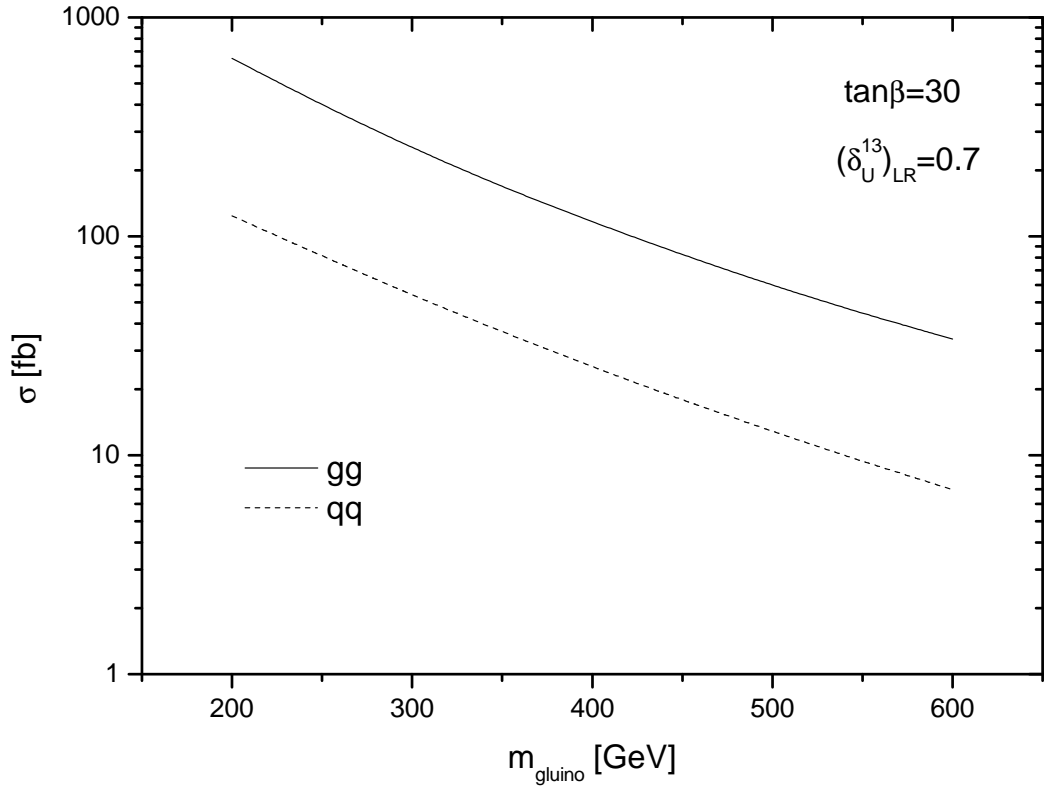, width=500pt}}
\caption[]{\label{fig:mgl}The total cross sections for the $pp \to
t \bar u $ as a function of the mass of gluino.}
\end{figure}

\begin{figure}
\centerline{\epsfig{file=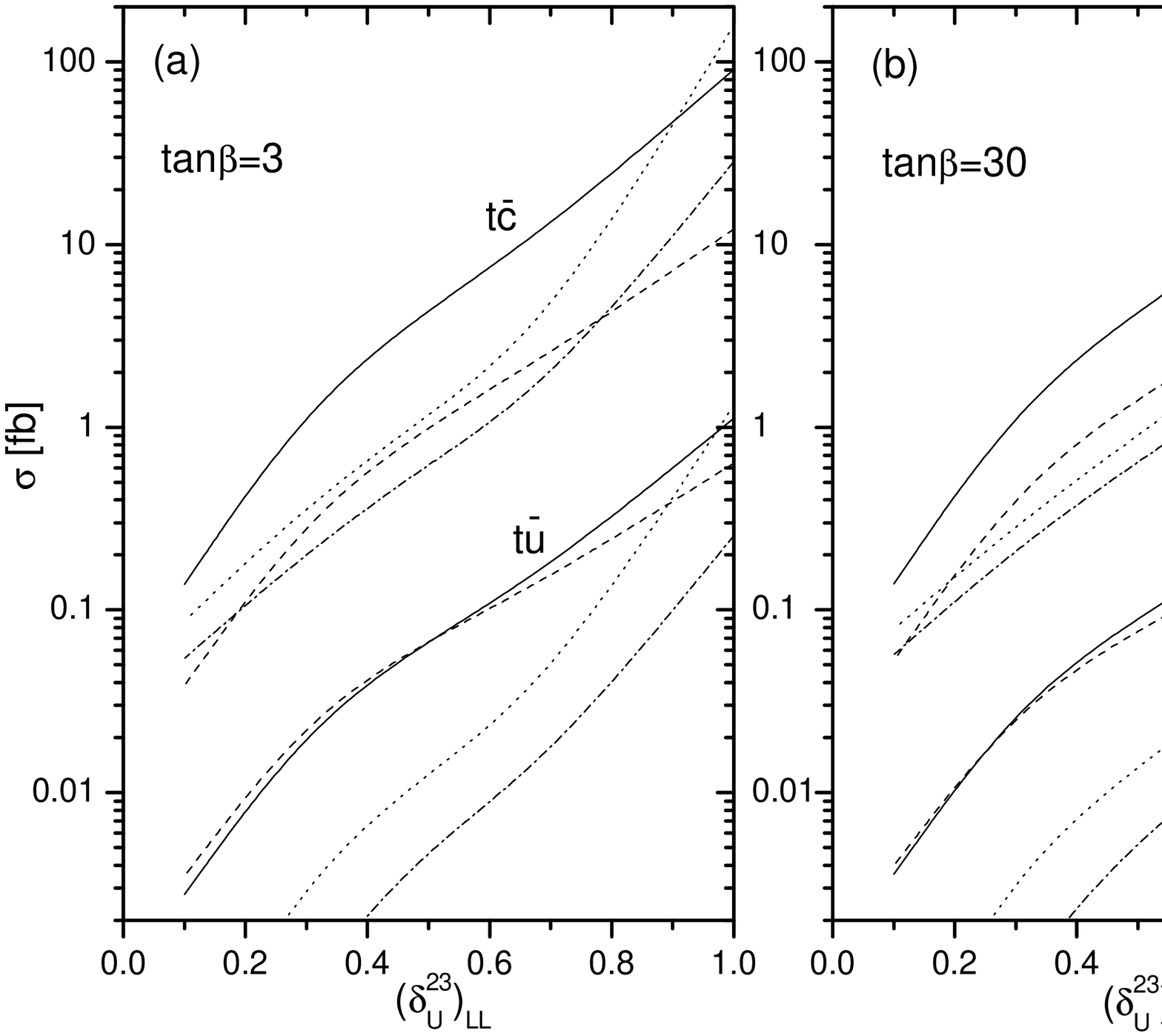, width=500pt}}
\caption[]{The total cross sections for the $pp \to t \bar c(\bar
u) $ with LL off-diagonal elements. Here, solid line:
$m_{\tilde{g}}=200$ GeV, $M_{\rm SUSY}=400$ GeV; dashed line:
$m_{\tilde{g}}=300$ GeV, $M_{\rm SUSY}=400$ GeV; dotted line:
$m_{\tilde{g}}=200$ GeV, $M_{\rm SUSY}=1000$ GeV; dash-dotted
line: $m_{\tilde{g}}=300$ GeV, $M_{\rm SUSY}=1000$ GeV.}
\end{figure}

\begin{figure}
\centerline{\epsfig{file=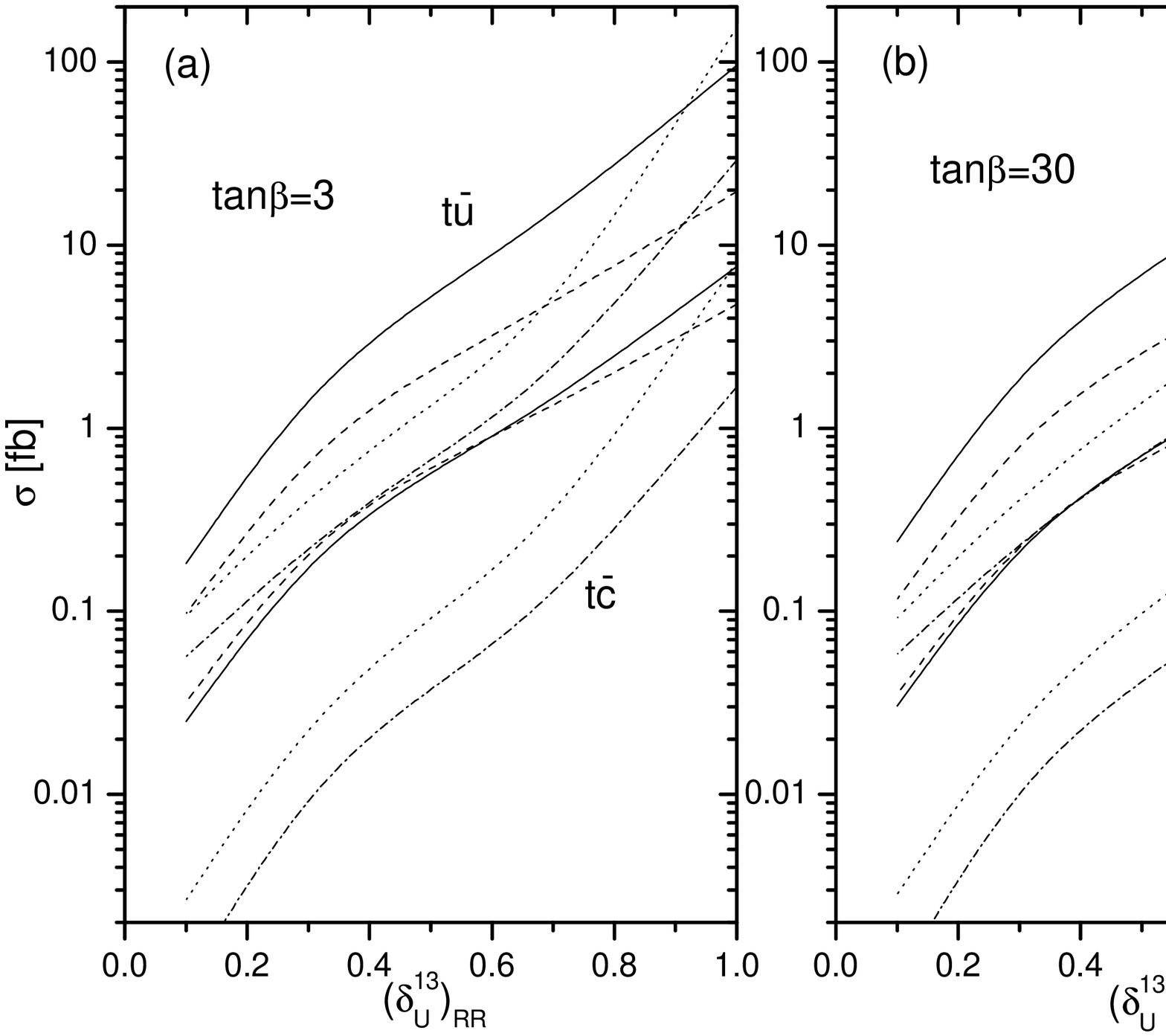, width=500pt}}
\caption[]{The total cross sections for the $pp \to t \bar c(\bar
u) $ with RR off-diagonal elements $(\delta^{13}_{U})_{RR}$. Here,
solid line: $m_{\tilde{g}}=200$ GeV, $M_{\rm SUSY}=400$ GeV;
dashed line: $m_{\tilde{g}}=300$ GeV, $M_{\rm SUSY}=400$ GeV;
dotted line: $m_{\tilde{g}}=200$ GeV, $M_{\rm SUSY}=1000$ GeV;
dash-dotted line: $m_{\tilde{g}}=300$ GeV, $M_{\rm SUSY}=1000$
GeV.}
\end{figure}

\begin{figure}
\centerline{\epsfig{file=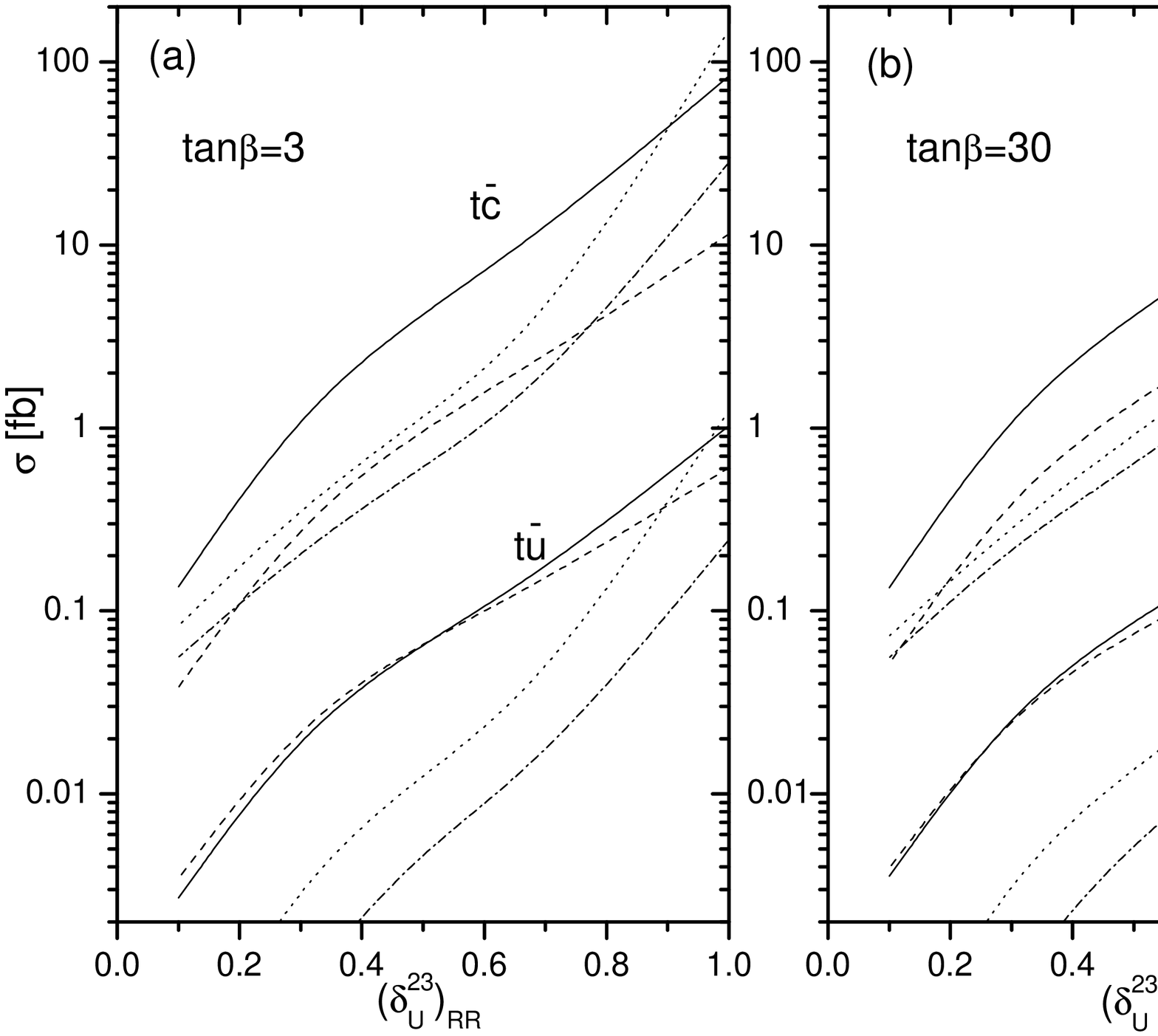, width=500pt}}
\caption[]{The total cross sections for the $pp \to t \bar c(\bar
u) $ with RR off-diagonal elements $(\delta^{23}_{U})_{RR}$. Here,
solid line: $m_{\tilde{g}}=200$ GeV, $M_{\rm SUSY}=400$ GeV;
dashed line: $m_{\tilde{g}}=300$ GeV, $M_{\rm SUSY}=400$ GeV;
dotted line: $m_{\tilde{g}}=200$ GeV, $M_{\rm SUSY}=1000$ GeV;
dash-dotted line: $m_{\tilde{g}}=300$ GeV, $M_{\rm SUSY}=1000$
GeV.}
\end{figure}

\begin{figure}
\centerline{\epsfig{file=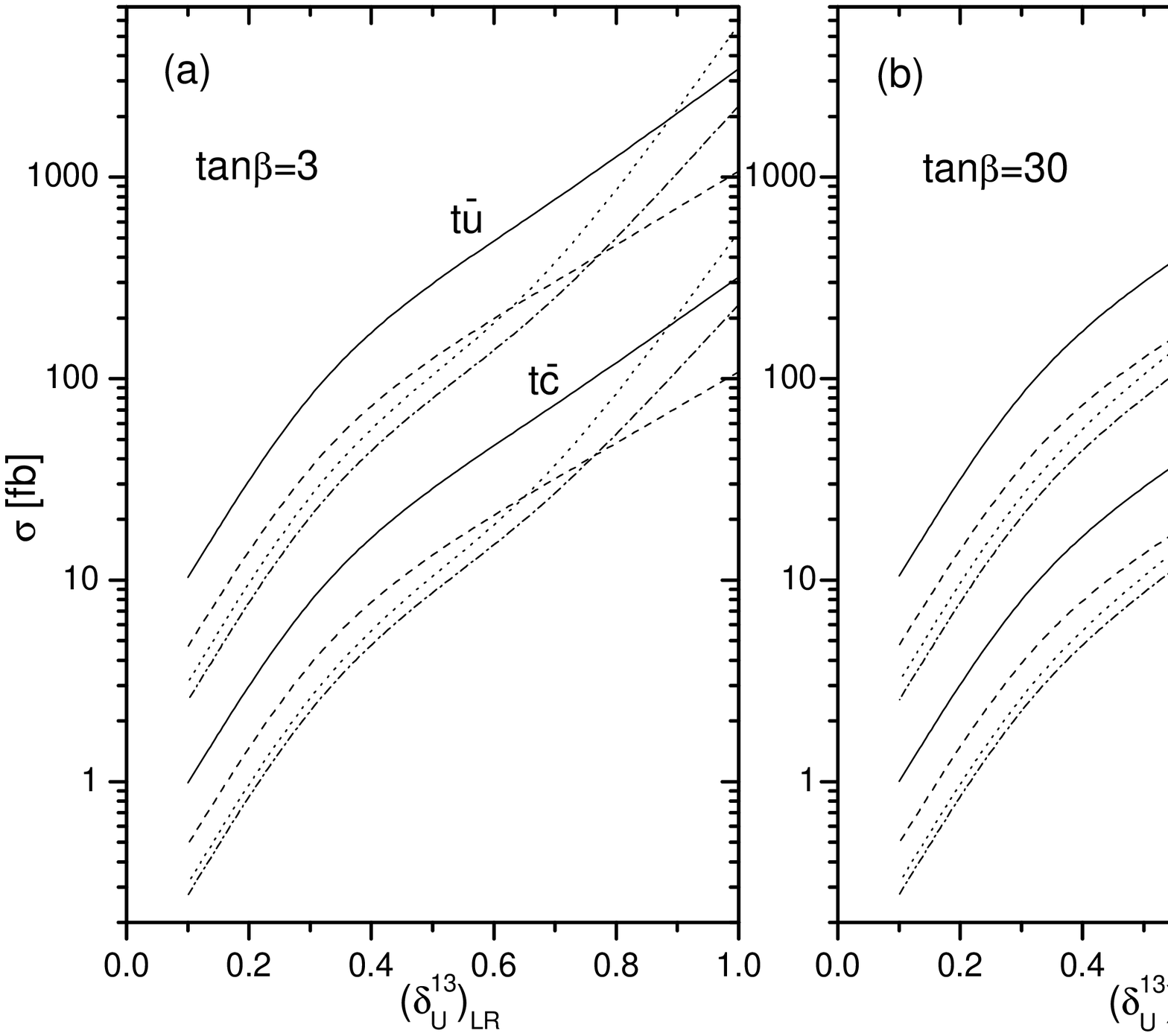, width=500pt}}
\caption[]{The total cross sections for the $pp \to t \bar c(\bar
u) $ with LR off-diagonal elements $(\delta^{13}_{U})_{LR}$. Here,
solid line: $m_{\tilde{g}}=200$ GeV, $M_{\rm SUSY}=400$ GeV;
dashed line: $m_{\tilde{g}}=300$ GeV, $M_{\rm SUSY}=400$ GeV;
dotted line: $m_{\tilde{g}}=200$ GeV, $M_{\rm SUSY}=1000$ GeV;
dash-dotted line: $m_{\tilde{g}}=300$ GeV, $M_{\rm SUSY}=1000$
GeV.}
\end{figure}

\begin{figure}
\centerline{\epsfig{file=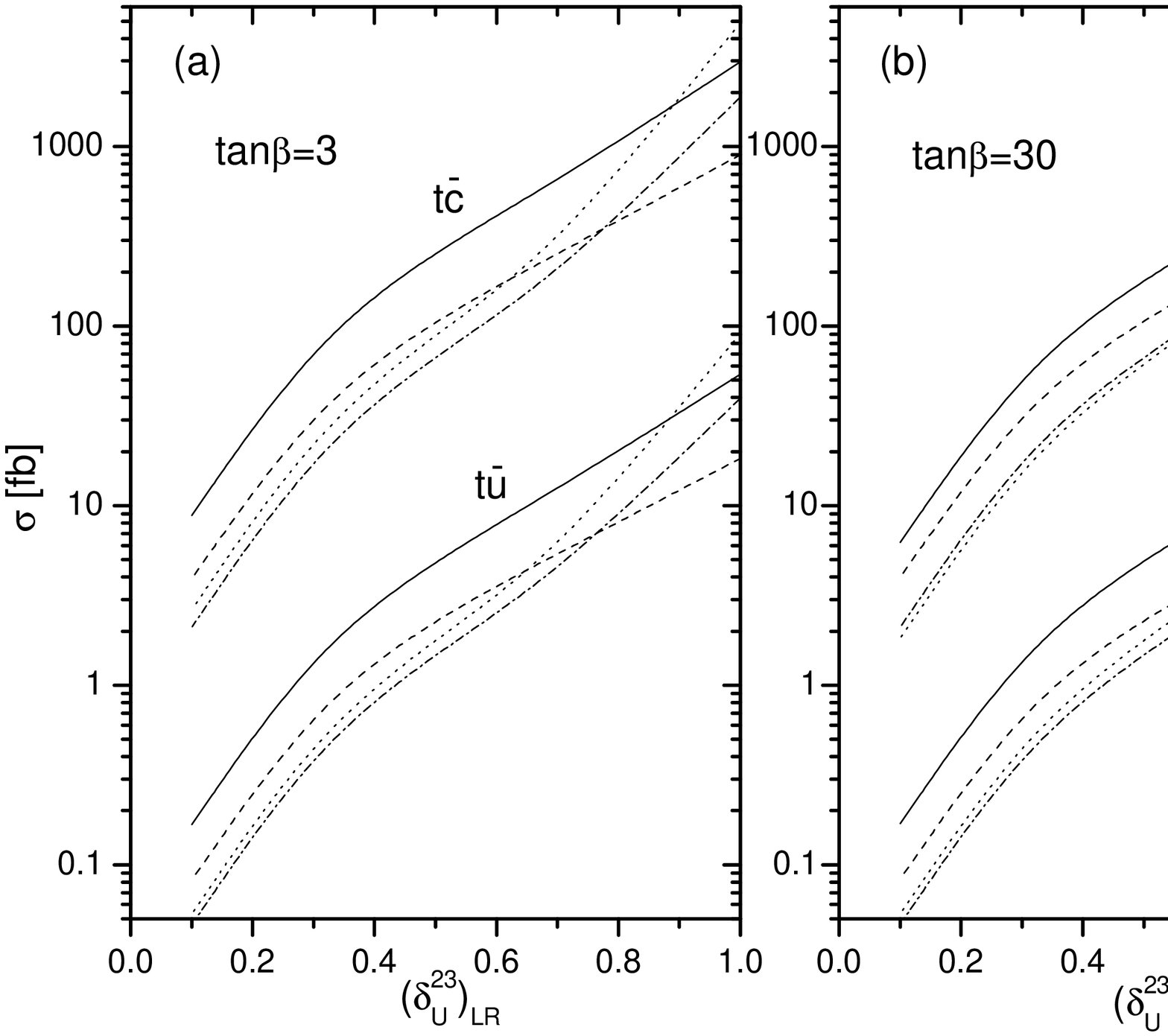, width=500pt}}
\caption[]{The total cross sections for the $pp \to t \bar c(\bar
u) $ with LR off-diagonal elements $(\delta^{23}_{U})_{LR}$. Here,
solid line: $m_{\tilde{g}}=200$ GeV, $M_{\rm SUSY}=400$ GeV;
dashed line: $m_{\tilde{g}}=300$ GeV, $M_{\rm SUSY}=400$ GeV;
dotted line: $m_{\tilde{g}}=200$ GeV, $M_{\rm SUSY}=1000$ GeV;
dash-dotted line: $m_{\tilde{g}}=300$ GeV, $M_{\rm SUSY}=1000$
GeV.}
\end{figure}

\begin{figure}
\centerline{\epsfig{file=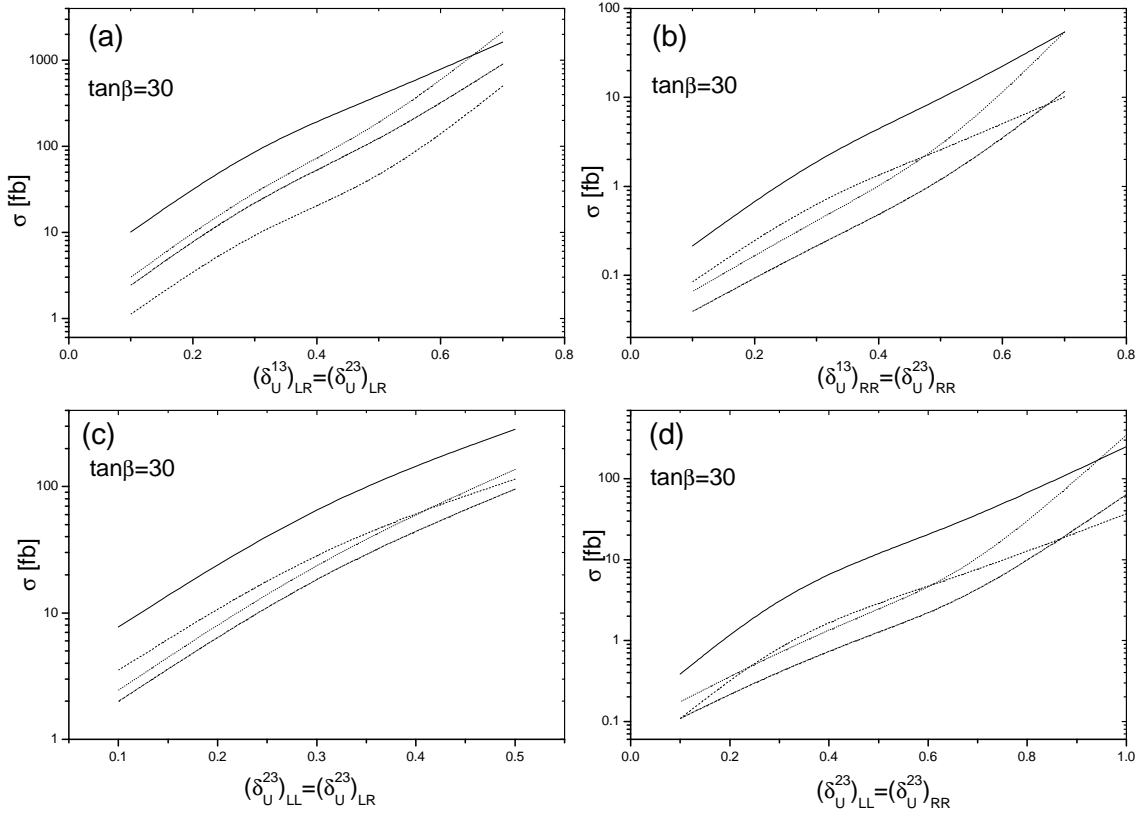, width=500pt}} \caption[]{Typical
interference effects between different matrix elements within one
block in (a) and (b), and  between different blocks in (c) and
(d). Here, solid line: $m_{\tilde{g}}=200$ GeV, $M_{\rm SUSY}=400$
GeV; dashed line: $m_{\tilde{g}}=300$ GeV, $M_{\rm SUSY}=400$ GeV;
dotted line: $m_{\tilde{g}}=200$ GeV, $M_{\rm SUSY}=1000$ GeV;
dash-dotted line: $m_{\tilde{g}}=300$ GeV, $M_{\rm SUSY}=1000$
GeV.}
\end{figure}

\begin{figure}
\centerline{\epsfig{file=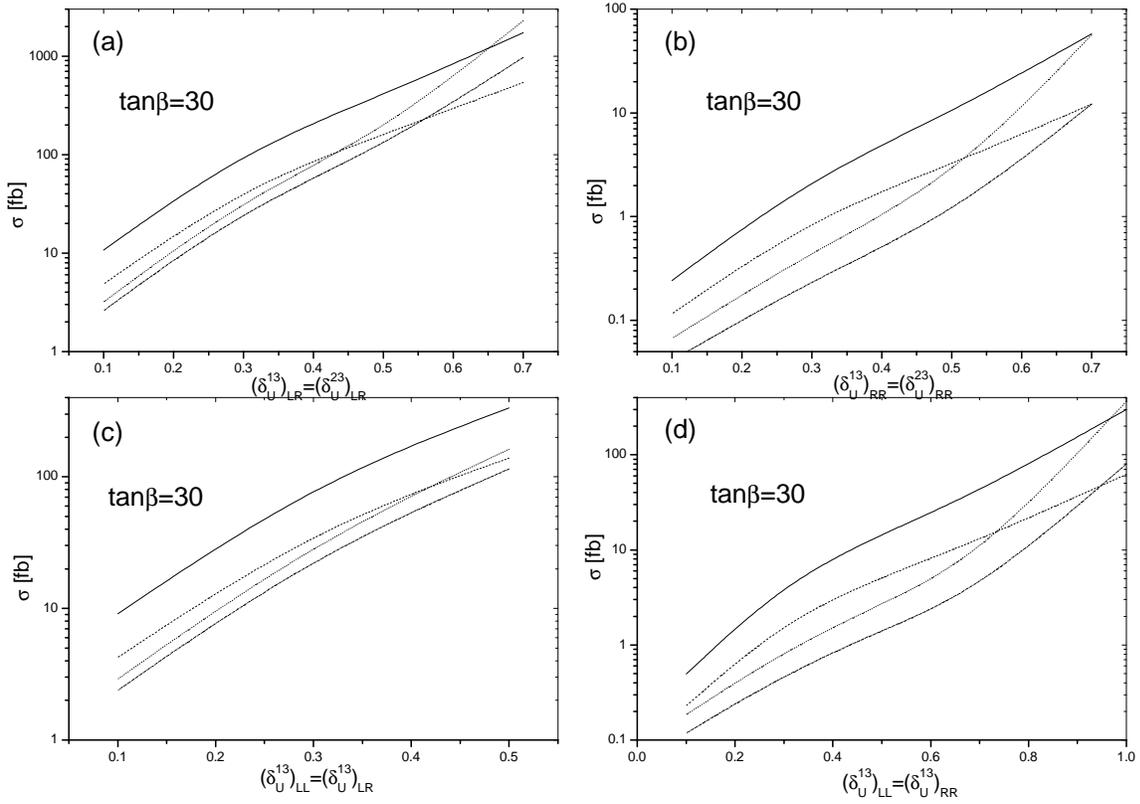, width=500pt}}
\caption[]{\label{fig:tumix}Similar as Fig.9, but for $t \bar u
$.}
\end{figure}

\newpage

\end{document}